\documentclass[12pt]{article}
\usepackage{amsmath,amsfonts,amssymb}
\usepackage{hyperref}
\usepackage{graphicx}
\usepackage[dvips]{color}
\usepackage{epsfig}
\usepackage{psfrag}
\usepackage{color}

\newcommand \be{\begin{eqnarray}}
\newcommand \ee{\end{eqnarray}}

\makeatletter\@addtoreset{equation}{section}\makeatother

\DeclareMathOperator{\Tr}{Tr}

\DeclareMathOperator{\sTr}{sTr}
\DeclareMathOperator{\Li}{Li}
\DeclareMathOperator{\Res}{Res}
\DeclareMathOperator{\arccot}{arccot}

\DeclareMathOperator{\sn}{sn}
\DeclareMathOperator{\cn}{cn}
\DeclareMathOperator{\dn}{dn}
\DeclareMathOperator{\am}{am}

\DeclareMathOperator{\sc1}{sc}
\DeclareMathOperator{\cd}{cd}

\def\bC {\mathbb{C}}
\def\bG {\mathbb{G}}

\def\bP {\mathbb{P}}
\def\bR {\mathbb{R}}
\def\bS {\mathbb{S}}
\def\bT {\mathbb{T}}

\def \bK {{\mathbb{K}}}

\newcommand{\gD}{{\boldsymbol D}}
\newcommand{\gG}{{\boldsymbol G}}
\newcommand{\gK}{{\boldsymbol K}}
\newcommand{\gL}{{\boldsymbol L}}
\newcommand{\gP}{{\boldsymbol P}}
\newcommand{\gQ}{{\boldsymbol Q}}
\newcommand{\gR}{{\boldsymbol R}}
\newcommand{\gS}{{\boldsymbol S}}

\newcommand{\beq}{\begin{equation}}
\newcommand{\eeq}{\end{equation}}
\newcommand{\bal}{\begin{equation}\begin{aligned}}
\newcommand{\eal}{\end{aligned} \end{equation}}
\newcommand{\bea}{\begin{eqnarray}}
\newcommand{\eea}{\end{eqnarray}}

\newcommand{\vev}[1]{{\left< {#1} \right>}}
\newcommand{\bra}[1]{{\left< {#1} \right|}}
\newcommand{\ket}[1]{{\left| {#1} \right>}}

\newcommand{\eqn}[1]{(\ref{#1})}

\newcommand{\address}[1]{\vbox{\center\em#1}}
\renewcommand{\title}[1]{\vbox{\center\huge{#1}}\vspace{5mm}}

\newcommand{\cB}{{\mathcal B}}
\newcommand{\cC}{{\mathcal C}}

\newcommand{\cN}{{\mathcal N}}
\newcommand{\cP}{{\mathcal P}}

\newcommand{\cO}{{\mathcal O}}

\newcommand{\psu}{{\mathfrak{psu}}}
\newcommand{\osp}{{\mathfrak{osp}}}

\begin{document}
\bibliographystyle{utphys}

\begin{titlepage}
\begin{center}
\phantom{ }

\vspace{15mm}

\title{Integrable Wilson loops}
\vspace{5mm}

\renewcommand{\thefootnote}{$\alph{footnote}$}

Nadav Drukker
\vskip 5mm
\address{
Department of Mathematics, King's College London \\
The Strand, WC2R 2LS, London, UK\\
\href{mailto:nadav.drukker@gmail.com}
{\tt nadav.drukker@gmail.com}}

\renewcommand{\thefootnote}{\arabic{footnote}}
\setcounter{footnote}{0}

\end{center}

\vskip5mm

\abstract{
\normalsize
\noindent
The generalized quark--antiquark potential of $\cN=4$ supersymmetric 
Yang-Mills theory on $\bS^3\times\bR$ calculates the potential between a pair of heavy 
charged particles separated by an arbitrary angle on $\bS^3$ and also an angle in flavor 
space. It can be calculated by a Wilson loop following a prescribed path and 
couplings, or after a conformal transformation, by a cusped Wilson loop in flat space, 
hence also generalizing the usual concept of the cusp anomalous dimension. 
In $AdS_5\times\bS^5$ this is calculated by an infinite open string. 
I present here an open spin--chain model which calculates the spectrum of excitations 
of such open strings. In the dual gauge theory these are cusped Wilson 
loops with extra operator insertions at the cusp. The boundaries of the 
spin--chain introduce a non-trivial reflection phase and break 
the bulk symmetry down to a single copy of $\psu(2|2)$. 
The dependence on the two angles is captured 
by the two embeddings of this algebra into $\psu(2|2)^2$, 
{\em i.e.}, by a global rotation. The exact answer to this problem is conjectured to be given by 
solutions to a set of twisted boundary thermodynamic Bethe ansatz integral equations. 
In particular the generalized quark-antiquark potential or cusp anomalous dimension 
is recovered by calculating the ground state energy of the minimal length 
spin--chain, with no sites. It gets contributions only from virtual particles 
reflecting off the boundaries. I reproduce from this calculation some 
known weak coupling perturtbative results.
}

\end{titlepage}


\begin{figure}[t]
\begin{center}
\begin{tabular}{ccc}
\raisebox{15mm}{\epsfig{file=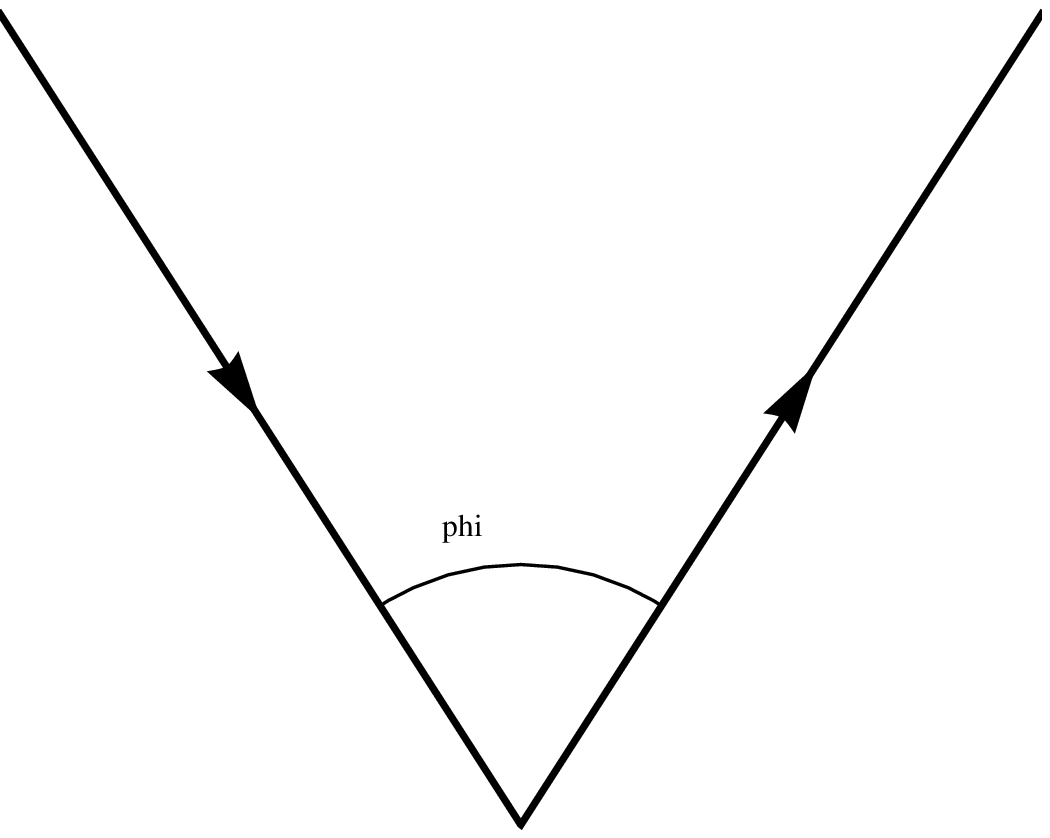,width=60mm
\psfrag{phi}{$\pi-\phi$}}} 
&\epsfig{file=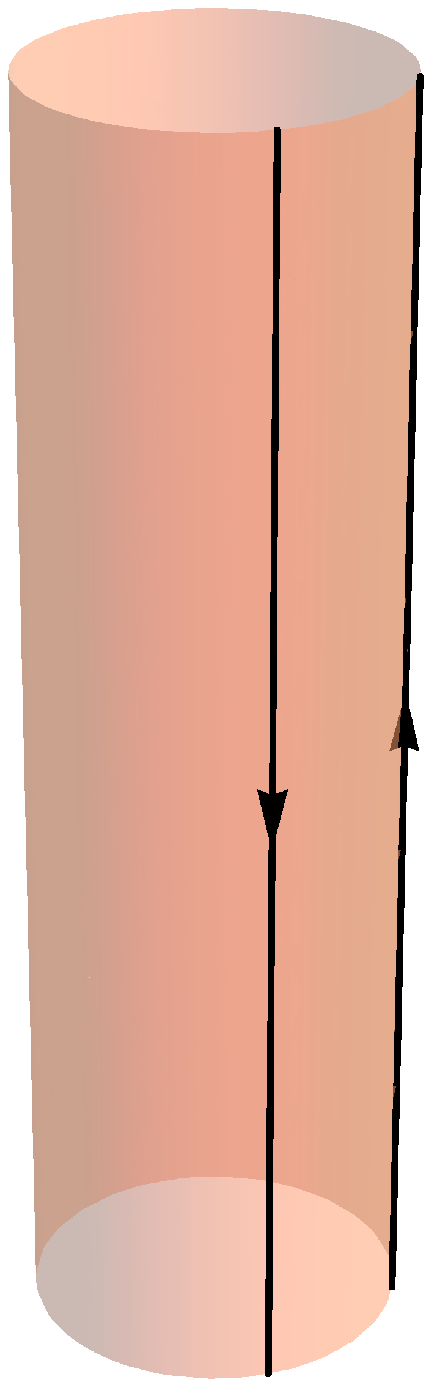,width=40mm} 
&\epsfig{file=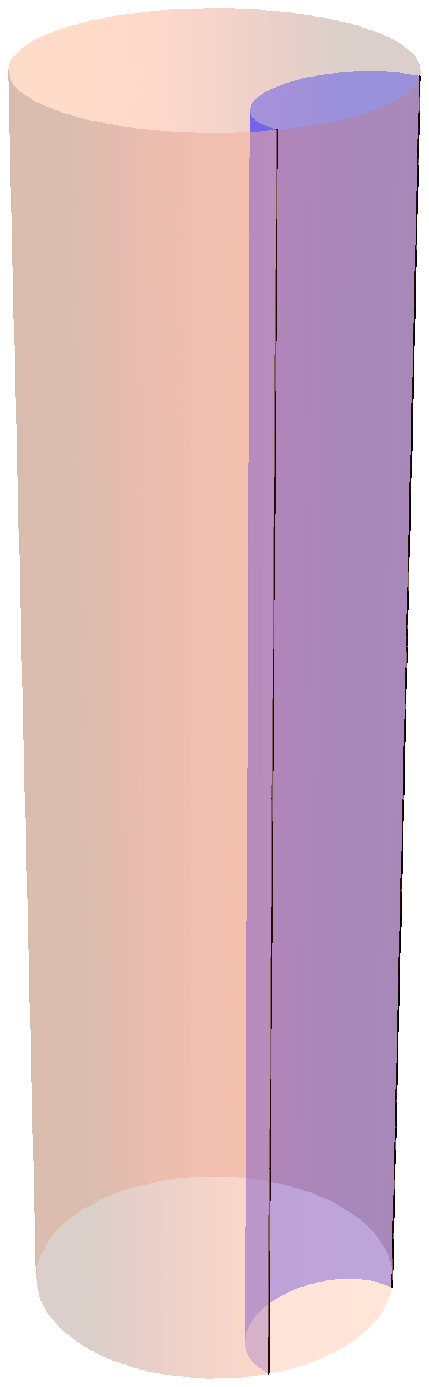,width=40mm} 
\\[-3mm]
$a$&$b$&$c$
\end{tabular}
\parbox{15cm}{
\caption{A cusped Wilson loop ($a$) in $\cN=4$ SYM on $\bR^4$ 
(and by analytic continuation also on $\bR^{3,1}$) is related to 
a pair of antiparallel lines ($b$) on $\bS^3\times\bR$. In the dual 
string theory this is calculated by a string world--sheet in 
$AdS_5\times S^5$ ending along the two lines on the boundary ($c$).
\label{fig:cylinder}}}
\end{center}
\end{figure}

\section{Introduction}
\label{sec:intro}

Wilson loops are some of the most important observables in nonabelian gauge theories. 
Among their many features, they capture the potential between heavy probe particles, 
hence can serve as an order parameter for confinement \cite{wilson}. 
They calculate a big part of the 
effect of high--energy scattering of charged particles 
\cite{Korchemsky:1988si,Korchemsky:1992xv}, and in the case of 
$\cN=4$ supersymmetric Yang-Mills theory (SYM) 
are conjectured to calculate scattering amplitudes exactly \cite{Bern:2005iz,am1}.

In \cite{Dru-For} a family of Wilson loop operators in $\cN=4$ SYM 
were presented and studied. Two useful points of view on these operators are 
(see Figure~\ref{fig:cylinder})
\begin{itemize}
\item
A cusp of angle $\pi-\phi$ in $\bR^4$, {\em i.e.}, two rays meeting at a point.
\item
Two lines along the time direction on $\bS^3\times\bR$, separated by an 
angle $\pi-\phi$ on the sphere. 
\end{itemize}
These two configurations are related to each other by a conformal transformation, so are 
essentially equivalent. The Maldacena-Wilson loops of $\cN=4$ SYM also include 
a coupling to a real scalar field and it is natural to allow the two rays/lines to couple to two different 
scalars, say $\Phi^1$ and $\Phi^1\cos\theta+\Phi^2\sin\theta$. Thus giving a two parameter 
family of observables, with the opening angle $\pi-\phi$ (such that $\phi=0$ is the straight 
line) and $\theta$.

These geometries calculate the physical quantities mentioned above. The pair of antiparallel 
lines gives the potential $V(\lambda,\phi,\theta)$ between two heavy charged particles 
propagating over a large time $T$ on $\bS^3$
\beq
\label{V}
\vev{W_\text{lines}}=e^{-T V(\lambda,\phi,\theta)}\,.
\eeq
It is possible to recover the potential in flat space by taking residue of $V$ as 
$\phi\to\pi$ (see \cite{Dru-For,CHMS2} for a detailed discussion). Cusped Wilson loops have the 
structure
\beq
\label{G}
\vev{W_\text{cusp}}=e^{-\Gamma_\text{cusp}(\lambda,\phi,\theta)\log\frac{\Lambda}{\epsilon}}\,,
\eeq
with the same function $\Gamma_\text{cusp}(\lambda,\phi,\theta)=V(\lambda,\phi,\theta)$ 
and $\Lambda$ and $\epsilon$ are IR and UV cutoffs, respectively. Scattering amplitudes are 
related to cusped Wilson loops in Minkowski space, so one should take $\phi=i\varphi$ to 
be imaginary. In particular in the limit of $\varphi\to\infty$, leads to 
$\gamma_\text{cusp}$ --- the {\em universal cusp anomalous dimension}
\beq
\Gamma_\text{cusp}(\lambda,i\varphi,\theta)\to \frac{\varphi\,\gamma_\text{cusp}(\lambda)}{4}\,.
\eeq

As expalined in detail in \cite{CHMS1}, in the limit of small $\phi$ and vanishing $\theta$, 
$\Gamma_\text{cusp}(\lambda,\phi,0)\sim -B(\lambda)\phi^2$ 
calculates also the radiation of a particle moving 
along an arbitrary smooth path. An exact expression is given there for $B$ 
including $1/N$ corrections to all orders in the gauge coupling $\lambda$. 
A modification of this expression applies also in expanding $\Gamma_\text{cusp}$ 
around the BPS configurations $\phi=\pm\theta$ \cite{zarembo}. In this manuscript an integrable 
spin--chain model is presented which is conjectured to calculate the full function $\Gamma_\text{cusp}$ 
in the planar approximation.

There is another application of this quantity, where it was originally defined 
\cite{Polyakov:1980ca,Brandt:1981kf}. Wilson loops satisfy a set of equations 
known as the loop equations \cite{Makeenko:1979pb}. 
These equations are nontrivial only for intersecting loops, 
where cusps appear. The loop equations have been solved in zero dimensional 
matrix models and in two dimensional Yang-Mills. To define them properly 
in four dimensional theories requires understanding cusped Wilson loops and 
how to renormalize their divergences. Logarithmic divergences of Wilson loops 
(and in the case of the Maldacena-Wilson loops in $\cN=4$ SYM, all the divergences) 
come from cusps, or from insertions of local operators. As explained in the following, 
integrability supplies the 
answer to this question in $\cN=4$ SYM, where all the logarithmic 
divergences arising from either insertions or cusps are governed by the same 
spin--chain model (to be precise, a set of twisted boundary thermodynamic 
Bethe ansatz equations (TBTBA, BTBA, or TBA in the follwoing)).

\begin{figure}[t]
\begin{center}
\begin{tabular}{ccc}
\raisebox{15mm}{\epsfig{file=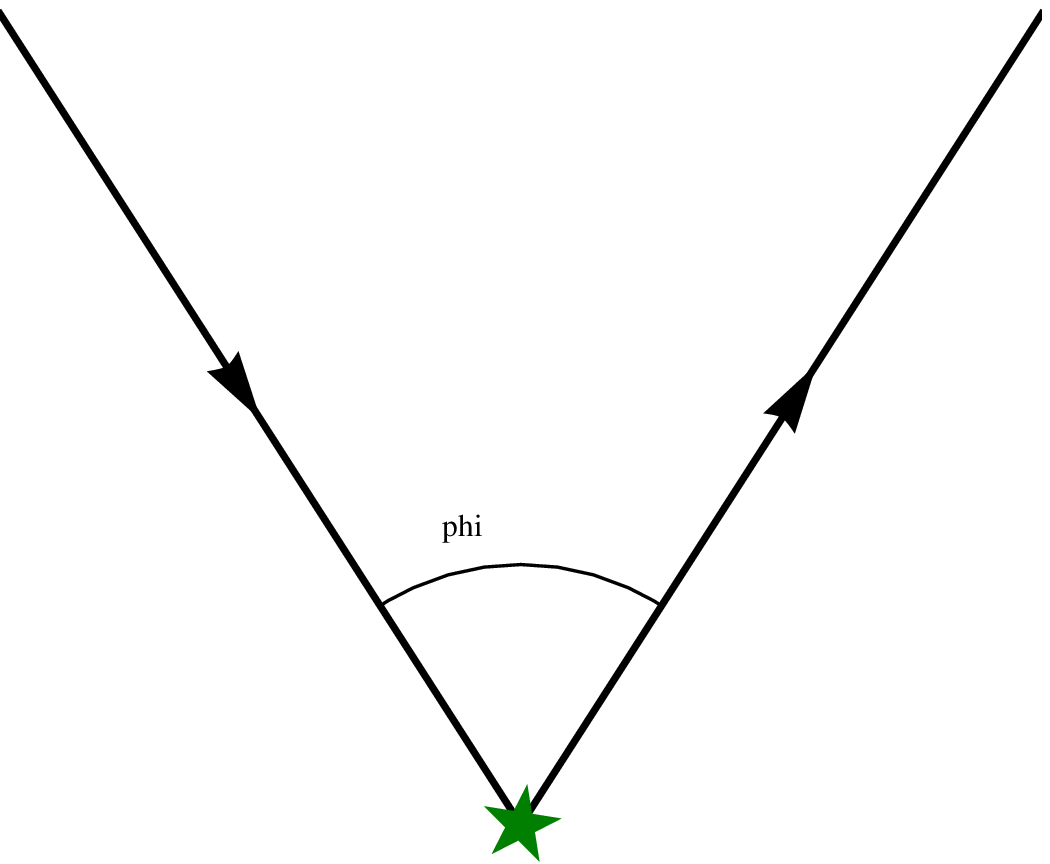,width=60mm
\psfrag{phi}{$\pi-\phi$}
\psfrag{i}{\raisebox{-7mm}{$\hskip-1.5cm\cO\sim ZYZ\cdots ZZ$}}
}} 
&\epsfig{file=cylinder1.eps,width=40mm}
\hskip-35mm
\raisebox{-8mm}[0mm][0mm]{\epsfig{file=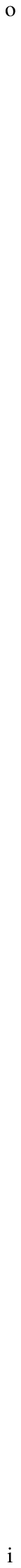,width=40mm
\psfrag{i}{\color [rgb]{0,0.3,0}$\,\ket{\psi}$}
\psfrag{o}{\raisebox{8mm}[0mm][0mm]{$\color [rgb]{0,0.3,0}\!\bra{\psi}$}}
}}
&
\hskip-7mm
\epsfig{file=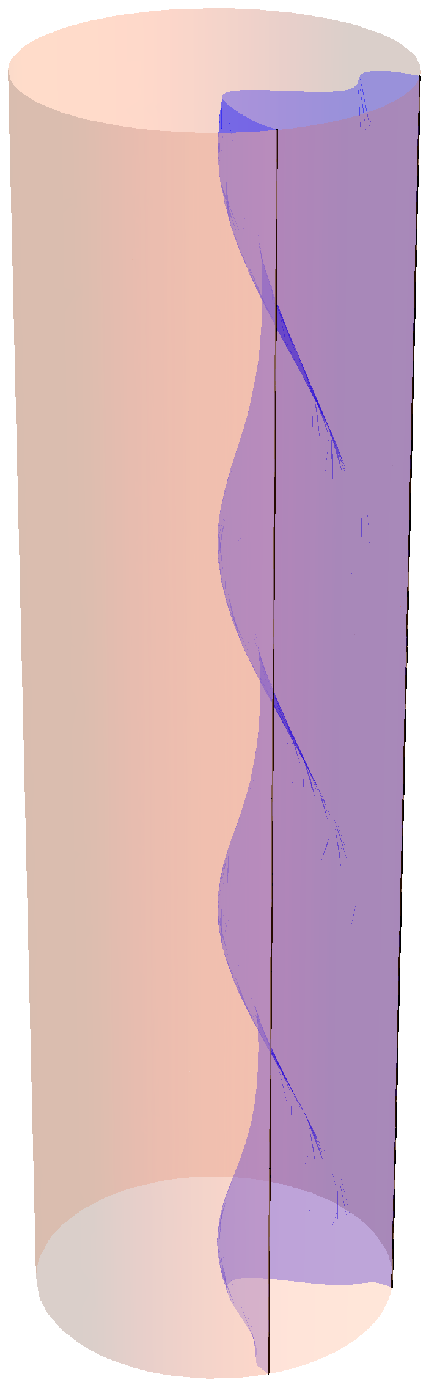,width=40mm} 
\\[-3mm]
$a$&$b$&$c$
\end{tabular}
\parbox{15cm}{
\caption{A generalization of Figure~\ref{fig:cylinder} 
allows an arbitrary adjoint valued local operator to be inserted at the apex 
of the cusp ($a$) in $\bR^4$. After the conformal transformation this is a 
pair of antiparallel lines ($b$) on $\bS^3\times\bR$, seemingly like in 
Figure~\ref{fig:cylinder}$b$, but in fact the details of the operator $\cO$ 
are represented by nontrivial boundary condition $\ket{\psi}$ 
at past and future infinity. The dual string solution in $AdS_5\times S^5$ 
still ends along the two lines, but is in an excited state 
(in the figure the sphere is suppressed).
\label{fig:excited}}}
\end{center}
\end{figure}

Integrability in this calculation is most easily seen in the string theory dual, where 
at leading order the Wilson loop is calculated by a classical string solution 
\cite{maldacena-wl,rey-yee,DGO,DGRT-big,Dru-For} and of course the string 
sigma--model in $AdS_5\times \bS^5$ is integrable \cite{Bena:2003wd}. 
At the one--loop 
level all the fluctuation operators about these particular classical solutions 
turn out to be versions of a known integrable 
system, the Lam\'e differential operator 
\cite{DGT,chr,vali-lines,Dru-For}. The calculations of the same quantity 
on the gauge theory side 
were based on brute force Feynman diagrammatics.

In order to see the integrability of these operators also in the gauge theory, 
it proves useful to consider 
a further generalization of the set of observables. 
Along a Wilson loop in the fundamental representation one can insert local operators 
in the adjoint representation of the gauge group while retaining gauge invariance. 
We shall insert such an operator at the apex of the cusp (Figure~\ref{fig:excited}). 
In the $\bS^3\times\bR$ picture this operator is at past 
infinity, and corresponds to an excitation of the Wilson loop.%
\footnote{The infinite lines are not well defined observables, as they are 
not invariant under gauge transformations at infinity. One should supply extra boundary 
conditions to construct them, and usually one assumes that the lines close onto each other 
in a smooth way. The excitations considered here have alternative boundary conditions.}

As mentioned, the expectation value of cusped Wilson loops suffer from logarithmic 
divergences, which can be considered the conformal dimension of the 
operator, meaning that these observable are eigenstates of the dilation 
operator. The generalized quark-antiquark potential $V$ in \eqn{V} (and 
$\Gamma_\text{cusp}$ in \eqn{G}) is 
exactly that conformal dimension. In order to understand this dimension as 
the solution of a spin--chain problem it is useful to consider the more general 
situation.

Note that in order to make this argument it is crucial that geometrically the 
cusp is invariant under dilations and the antiparallel lines under translations 
(also in addition a $U(1)$ of rotations in the transverse space and 
an $SO(4)$ flavor symmetry). 
Therefore one can define the problem of finding operators in the quantum 
theory which are eigenstates under this transformation. A similar argument 
cannot be implemented in a trivial way for Wilson loops of other shapes.

Another way to visualize the situation in the dual string theory, where the Wilson 
loops are described by a string stretching between the two lines at the boundary of 
$AdS_5\times\bS^5$. Thus, along the boundary of the open string we have two 
different boundary conditions, corresponding to the two lines. In world-sheet 
language we should consider the string with the insertion of two operators, 
these are not regular open string states, rather they are {\em boundary changing 
operators}. To be precise, for any pair of boundary conditions there is a 
family of such boundary changing operators exactly corresponding to the cusped 
Wilson loops with extra insertions. The original cusp without an extra operator 
insertion maps to the ground state of the open string, {\em i.e.}, the 
boundary changing operator of the lowest dimension.

The case of the insertion of adjoint operators into the straight Wilson line 
with $\theta=\phi=0$ was 
studied in \cite{dk-spinchain}. In the string theory dual these operators are mapped 
to regular open--string vertex operators as the boundaries of the string do not 
get modified. On the gauge theory side the dimension of these insertions can 
be calculated by solving an open spin--chain problem, as is reviewed in the next 
section and extend to the more general situation.

If the operators carry macroscopic amount of angular momentum in $AdS_5$ or 
$\bS^5$, then there is a semiclassical description for them in string theory. For the 
insertions into the straight line this was done in \cite{dk-spinchain}, and it can 
be generalized for an arbitrary cusp using the techniques in \cite{df-int}. These 
solutions are such that the string approaches the boundary along the same two lines 
as the solutions without the insertions (see Appendix~B of \cite{Dru-For}), 
but performing extra rotations around the $\bS^5$ or in $AdS_5$ in the bulk 
of the world-sheet. 

\begin{figure}[t]
\begin{center}
\epsfig{file=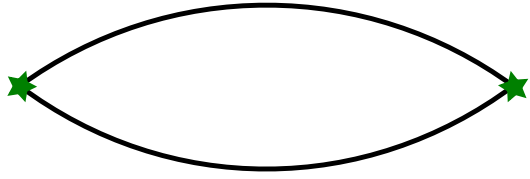,width=70mm}
\parbox{15cm}{
\caption{Yet another picture of a Wilson loop with two cusps 
(with possible local insertions) connected by arcs. 
It is related to that in Figure~\ref{fig:excited}$a$ by a conformal transformation 
mapping the point at infinity to finite distance. If the distance between the 
cusps is $d$, the expectation value of this Wilson loop is 
$\vev{W}\propto 1/d^{2V(\lambda,\phi,\theta)}$, where the logarithmic 
divergences in \eqn{G} are interpreted, as usual as renormalizing the 
classical dimension.
\label{fig:arcs}}}
\end{center}
\end{figure}

To be completely explicit, I write down the form of the Wilson loop operator with insertion 
of the simplest adjoint valued local operator $\cO=Z^J$ with 
$Z=\Phi^5+i\Phi^6$
\begin{align}
W&=\cP\,e^{\int_{-\infty}^0[iA_\mu(x(s))\dot x^\mu+\Phi^1(x(s)|\dot x|)ds}\,Z(0)^J
e^{\int_0^\infty(i A_\mu(x(s))\dot x^\mu
+\cos\theta\,\Phi^1(x(s))|\dot x|
+\sin\theta\,\Phi^2(x(s))|\dot x|]ds}
\nonumber\\
&=\cP\,e^{\int_{-\infty}^0[iA_1(x(s))+\Phi^1(x(s))ds}\,Z(0)^J
e^{\int_0^\infty(i\cos\phi\, A_1(x(s))
+i\sin\phi\, A_2(x(s))
+\cos\theta\,\Phi^1(x(s))
+\sin\theta\,\Phi^2(x(s))]ds}
\nonumber\\
x(s)&=\begin{cases}
(s,0,0,0)\quad&s\leq0\,,\\
(s\cos\phi,s\sin\phi,0,0)\quad&s\geq0\,,\\
\end{cases}
\end{align}
By a conformal transformation the point at infinity can be mapped to finite distance 
(see Figure~\ref{fig:arcs}) and then the Wilson loop is 
a completely kosher gauge invariant operator with two cusps with local operators 
$Z^J$ inserted at one and $\bar Z^J$ at the other.

$V(\lambda,\phi,\theta)$ was calculated in \cite{Dru-For} to second order 
in perturbation theory.%
\footnote{The three loop answer was derived very recently in \cite{CHMS2}.}
The result is
\begin{align}
V(\lambda,\phi,\theta)
&=-\frac{\lambda}{8\pi^2}\,\frac{\cos\theta-\cos\phi}{\sin\phi}\,\phi
\nonumber\\&\quad
+\left(\frac{\lambda}{8\pi^2}\right)^2
\bigg[\frac{1}{3}\,\frac{\cos\theta-\cos\phi}{\sin\phi}
\,(\pi^2-\phi^2)\phi
\label{pert}
\\
\nonumber
&\quad\qquad
-\frac{(\cos\theta-\cos\phi)^2}{\sin^2\phi}
\left(\Li_3\left(e^{2i\phi}\right)-\zeta(3)
-i\phi\left(\Li_2\left(e^{2i\phi}\right)+\frac{\pi^2}{6}\right)
+\frac{i}{3}\phi^3\right)\bigg],
\end{align}

In the next section the spin--chain model is presented. Section~\ref{sec:wrapping} 
calculates the leading weak coupling contribution to the the cusped Wilson loop 
from the exchange of a single virtual magnon (L\"uscher-Bajnok-Janik correction). 
Section~\ref{sec:BTBA} presents the twisted boundary TBA which calculates 
this quantity at all values of the coupling. Finally Section~\ref{sec:discuss} 
discusses some of the results and possible generalizations. Some technical details 
are relegated to appendices.

At an advanced state of this project I learnt of \cite{CMS}, which has a great deal 
of overlap with the results reported in this manuscript.

\section{Wilson loops and open spin--chains}
\label{sec:open-chains}

In this section the relation between the insertion of adjoint valued operators 
into a Wilson loop and open spin--chains is developed. First the main principles 
are explained in general 
terms and only later are the precise formulas for the open spin--chain presented.

In the case of gauge invariant local operators the spin--chain arises 
\cite{Minahan:2002ve} by choosing a reference ground--state $\Tr Z^J$ with 
$Z=\Phi^5+i\Phi^6$ (also known as the BMN vacuum \cite{bmn}) and 
considering the replacement of some of the $Z$ fields by other fields of the 
theory: The other scalars, the fermions or field--strenghts. 
In addition one can act with derivative operators. The possible insertions 
are labeled by representations of $PSU(2|2)^2\subset PSU(2,2|4)$ which is the residual 
symmetry preserved by the vacuum. These states are viewed as excitation of a spin--chain 
and it is conjectured that this spin--chain is integrable and satisfies a particular 
dispersion relation leading to a solution in principle of the spectral problem of the 
theory, which has passed many stringent tests. The symmetry 
is reviewed in Appendix~\ref{app:susy} and the scattering 
matrix of the fundamental representation of this spin--chain is presented in 
Appendix~\ref{app:magnons}.

The same thing can be done with insertions of adjoint valued local operators into 
Wilson loops, as explained some time ago in \cite{dk-spinchain}. That paper 
studied only the insertions of operators made of two complex scalar fields 
($Z$ and $Y=\Phi^4+i\Phi^5$ forming the $SU(2)$ sector) into a straight (or circular) 
Wilson loop. Here the construction is 
generalized to the full set of allowed states. Also, the crucial modification 
of the spin--chain model when replacing a straight line with an arbitrary 
cusp is explained.

\begin{figure}[t]
\begin{center}
\epsfig{file=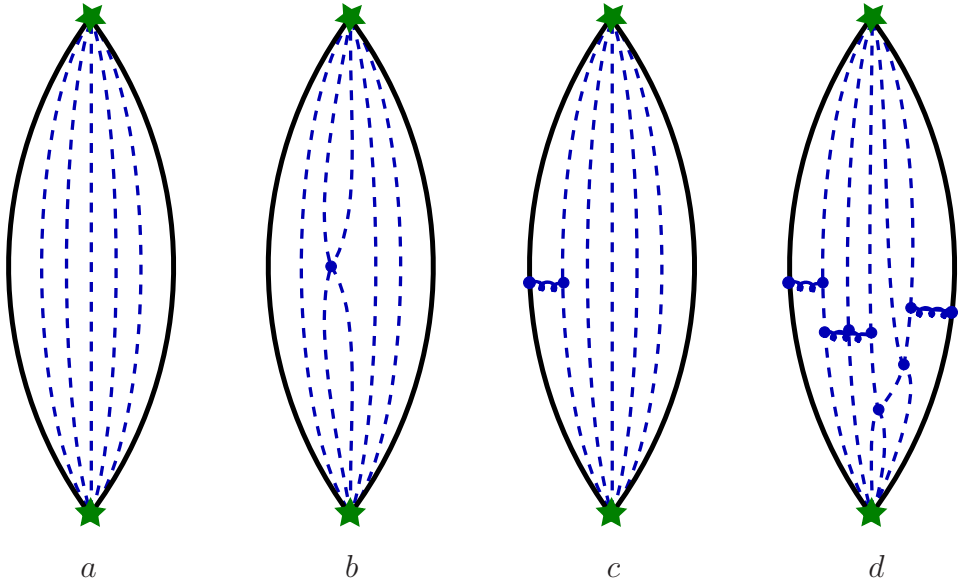,height=71mm}
\\
$a\hskip1.3in
b\hskip1.3in 
c\hskip1.3in
d$
\\
\parbox{15cm}{
\caption{Sample planar Feynman graph calculations for a Wilson loop with two 
insertion of operators made of five scalar fields at the cusps 
At leading order there are only five free propagators (dashed lines)
so the classical conformal dimension is five ($a$). 
At one loop there are nearest neighbor interactions among the 
scalar fields (like $b$), which are identical to those of single trace local operators, 
but also interaction between the last scalar and the Wilson loop ($c$). 
At this order planar graphs 
connecting the loop to itself are restricted to each of the arcs, they are 
finite and do not modify the conformal dimension. Finally The two arcs interact by
``wrapping effects'', which arise in this example only at six loop order ($d$). 
Up to this order the dimension of an operator made of the scalars $Z$ and $Y$ 
does not depend on the cusp angles $\phi$ and $\theta$.
\label{fig:graphs}}}
\end{center}
\end{figure}

The basic principle of \cite{dk-spinchain}, illustrated in Figure~\ref{fig:graphs}, 
is that the calculation of the Wilson loop with local insertions 
is very similar to that of the dimension of local operators. 
At one loop order in the planar approximation every field in the local insertion interacts 
directly only with its two nearest neighbors. Only the first and last 
fields interact with the Wilson loop itself. This leads naturally to an 
open spin--chain model, where the bulk hamiltonian is the same as the 
usual one as in the closed spin--chain, but with two boundaries where one 
must specify boundary conditions, {\em i.e.}, reflection matrices.

It is important to emphasize that this is based mainly on abstract arguments 
and on very few explicit calculations, which is also the case (but to a lesser 
extent) with the usual 
spectral problem. It is clear from diagramatics that indeed the bulk 
spin--chain hamiltonian is identical to the closed spin--chain one. The hamiltonian 
in any case is known explicitly only at low orders in perturbation theory, but 
the scattering matrix is known and therefore expected to be the same. The boundary 
interactions were calculated in \cite{dk-spinchain} only in the $SU(2)$ sector 
and only at one loop, where they are essentially trivial. Below, an exact expression 
for the reflection matrix is proposed, which should capture the boundary interactions 
at all loop order. It is clearly conjectural, based on the assumption of integrability, 
representation theory, 
boundary crossing symmetry and minimality of the dressing factor (see below). 
The formalism presented does reproduce the results of the perturbative calculation, 
which lends credence to this conjecture.

There are other manifestations of open spin--chains in $\cN=4$ SYM, which 
apart for the Wilson loops is whenever there are D-branes in the dual 
$AdS_5\times\bS^5$ space, on which the open strings can end.%
\footnote{The strings 
describing the Wilson loop are also open, but extend all the way to the 
boundary of space, rather than ending on a D-brane.}

The most studied example is that of the D3-brane giant graviton, which is 
a determinant operator in the field theory, see 
\cite{Berenstein:2005vf,Mann:2006rh,Hofman:2007xp}. 
A closely related system is that of $AdS_5$ filling 
D7-branes introducing fundamental matter into the theory which can serve 
as the ends of an adjoint valued word, again leading to open spin--chains.

Another system, that of D5-branes representing 3d defects in the gauge theory 
(domain walls with fundamental matter) \cite{DeWolfe:2004zt} 
has a symmetry much closer 
to that of the Wilson loops and has been erroneously thought for a long 
time not to lead to integrable boundary conditions on the spin--chain 
\cite{Correa:2008av}. This has been corrected recently in \cite{CRY,GMU} 
(see also \cite{Dekel:2011ja}), and is very useful for the derivation below.

Before going into the technical details, it is useful to explore some of the 
features that can be extracted from looking at the Feynman diagrams, as in 
Figure~\ref{fig:graphs}.

The modification from the straight line to the cusped loop (or from the circle to 
the pair of cusps as in Figure~\ref{fig:arcs}) is quite easy. The spin--chain 
has to satisfy two different boundary conditions at the two ends. 
The two rays of the Wilson loop 
couple to linear combinations of the scalars $\Phi^1$ and $\Phi^2$ 
break the $SO(6)$ $R$-symmetry to $SO(4)$. The ground state $Z^J$ 
breaks it further 
to $U(1)$ rotating $\Phi^3$ into $\Phi^4$. It also has charge $J$ 
under the $U(1)$ which rotates $\Phi^5$ into $\phi^6$.

The fields $Z$ and $Y$ (also $\bar Z$ and $\bar Y$) interact with the 
two segments of the Wilson loop in the same way, at the one loop order 
they have purely reflective boundary conditions, so the eigenstates 
of the one loop hamiltonian of the form $\sum_k e^{ip k}\,Z^{k-1}YZ^{J-k}$ 
are Neumann functions. The fields 
$\Phi^1$ and $\Phi^2$ (or $X$ and $\bar X$) interact differently 
with the two boundaries, a linear combination of them (depending on 
the scalar coupling of the Wilson loop) has to vanish at the boundary, 
so they satisfy partially Dirichlet and partially Neumann conditions and 
the allowed magnon momenta will depend on $\theta$.

Going to higher order in perturbation theory doesn't change much until 
order $J+1$. The 
range of interaction in the spin--chain is identical to the order in perturbation 
theory and at this order there are graphs which communicate all the way 
from one side to the other (Figure~\ref{fig:graphs}$d$). In analogy with 
the closed spin--chain these can be called ``wrapping corrections''. 
For operators made of $Z$ and $Y$ these 
are the first graphs where the answer will depend on the cusp 
angles $\phi$ and $\theta$. In particular the 
ground state $Z^J$ has dimension $\Delta=J$ up to 
order $\lambda^J$ and gets corrections {\em only} from wrapping 
effects (and only for $\phi\neq\theta$, when the system is not globally 
BPS \cite{zarembo}). The wrapping corrections are hard to 
calculate directly for large $J$, but in the spin--chain formalism they 
are given by L\"uscher like corrections \cite{Luscher:1985dn,bajnok-janik}, as discussed in 
Section~\ref{sec:wrapping}. Higher order wrapping effects are best 
captured by the twisted boundary TBA equations. 

Note though, that in the case of primary interest, that without the local 
insertion, $J=0$, so wrapping corrections contribute at one loop order, 
double wrapping at two loops and so on. In that case the diagramatics 
are not that hard and indeed these calculations were done up to three loop 
order in \cite{DGO,mos,Dru-For,CHMS2} (see also \cite{kr-wl,Kotikov:2003fb}). 
This system then provides an interesting laboratory to 
study the twisted boundary TBA equations, where the desired quantity 
is the ground state energy, which is the simplest observable in the TBA. 
This is discussed in detail in Section~\ref{sec:BTBA} below.

\subsection{Boundary symmetries}
\label{sec:bsusy}

It is now time to start with the detailed analysis of the system. The first question is 
how the symmetry of the theory and of the usual spin--chain model is modified by the 
presence of the boundaries. It is crucial that one can focus on one boundary at a time, 
since the symmetry preserved by each boundary is significantly larger and more 
restrictive than that preserved by both together.

The $\psu(2,2|4)$ supersymmetry algebra of the theory is written down in 
Appendix~\ref{app:susy} as is its breaking 
to $\psu(2|2)^2\times\mathfrak{u}(1)$ by the choice of ground state. 
Here we describe the breaking of the symmetry by the Wilson loop, which 
furnishes the spin--chain with boundaries.

A straight Wilson loop in the $x^\mu$ direction and coupling to the scalar $\Phi^I$ 
preserves half the supercharges, those given by the combinations
\beq
\label{WLsusy}
Q^\alpha_A+\epsilon^{\alpha\beta}\gamma^\mu_{\beta\dot\gamma}\rho^I_{AB}\bar Q^{\dot\gamma B}\,,
\qquad
\bar S_{\dot\alpha A}-\epsilon_{\dot\alpha\dot\beta}\gamma^{\mu\,\dot\beta\gamma}\rho^I_{AB}S^B_\gamma\,.
\eeq
Here $\gamma^\mu$ are the usual gamma matrices in space-time and $\rho_I$ are 
those for the $SO(6)$ flavor symmetry.

These generators close onto an $\osp(4^*|4)$ algebra. It is possible to include an extra 
phase $e^{i\alpha}$ between the two terms in both sums and the algebra still closes 
to an isomorphic algebra. For $\alpha=\pi/2$ this is the symmetry of an 't~Hooft loop and 
other values correspond to dyonic loop operators. So this is the symmetry preserved 
by each of the boundaries.

Consider the Wilson loop with $\mu=I=1$ and the choice of $SO(6)$ gamma matrices
\bal
&\rho_1^{14}=\rho_1^{23}=1\,,
&\qquad
&{-}\rho_3^{13}=\rho_3^{24}=1\,,
&\qquad
&\rho_5^{34}=\rho_5^{12}=1\,,
\\
&\rho_2^{14}=-\rho_2^{23}=i\,,
&\qquad
&\rho_4^{13}=\rho_4^{24}=i\,,
&\qquad
&\rho_6^{34}=-\rho_6^{12}=i\,.
\eal
Of the two copies of $\psu(2|2)$ annihilating the vacuum $Z^J$ \eqn{su222}, only one diagonal copy 
survives, once considering the linear combinations \eqn{WLsusy} annihilating the 
Wilson loop. Those are
\beq
Q_a^\alpha+i(\sigma^3)^\alpha_{\ \dot\beta}(\sigma^3)^{\dot b}_{\ a}\bar Q_{\dot b}^{\dot\beta}\,,
\qquad
\bar S^{\dot a}_{\dot\alpha}-i(\sigma^3)^\beta_{\ \dot\alpha}(\sigma^3)^{\dot a}_{\ b}S^b_\gamma\,.
\label{diag}
\eeq

A similar algebra, though with a different real form and embedding into $\psu(2,2|4)$, 
namely $\osp(4|4,\bR)$, is also preserved by defects represented in 
$AdS_5\times\bS^5$ by D5-branes. 
As is shown in \cite{CRY} the open spin--chain with such boundary conditions 
has the same boundary symmetry which is the intersection 
$\osp(4|4,\bR)\cap\psu(2|2)^2=\psu(2|2)$.

In that case the defect breaks the $SO(6)$ $R$ symmetry to $SO(3)\times SO(3)$, while 
for the Wilson loop it is broken to $SO(5)$. In that case (as in other realizations 
of open spin--chains in $\cN=4$ SYM) there are two choices for the vacuum, depending 
on which copy of the two inequivalent $SO(3)$ is broken by the choice of ground state. 
In one case the resulting boundary is charged under the unbroken $SO(3)$, and therefore 
carries a representation of $\psu(2|2)$. In the other case, it is uncharged and does not 
carry a boundary degree of freedom. In the case of the Wilson loop it is natural to 
break the $SO(5)$ symmetry to $SO(3)$ and the resulting boundary has no degrees 
of freedom.
This can be seen by explicit calculation, since the straight Wilson loop with 
an insertion of $Z^J$ is a perfectly good BPS operator and does not require other 
fields ``shielding'' the local operator from the Wilson loop (as would be the case with 
a vacuum like $X^J$ with $X=\Phi^1+i\Phi^2$).%
\footnote{The $X^J$ state breaks $SO(5)\to SO(4)$ and would require boundary 
excitations. But it is not a good vacuum, as can be seen by the fact that it does not 
share any supersymmetry with the  Wilson loop.}
This can also be seen from the dual string point of view, where the description of the 
string ground state \cite{dk-spinchain} does not require breaking the symmetry and 
the resulting Goldstone bosons of global rotation in $AdS_5\times\bS^5$. The 
spin--chain calculation below can be quite easily generalized to the case of 
boundary degrees of freedom to describe both type of D5-brane open spin--chains.

\subsection{Notations}
\label{sec:notations}

Before writing down the spin--chain model which calculates these Wilson loop 
operators let us fix some notations.

The magnons can be characterized by the spectral 
parameters $x^\pm$. Introducing the parameter $u$, then 
for a general bound state of $Q$ magnons in the physical domain they satisfy
\beq
x^\pm+\frac{1}{x^\pm}=\frac{u}{g}\pm\frac{iQ}{2g}\,,
\qquad
g=\frac{\sqrt\lambda}{4\pi}\,.
\eeq
For generic values of $u\in\bC$ there are four possible solutions to the above 
equations with $x^+$ and $x^-$ both outside the unit disk, both inside and 
with either one outside and the other inside.
 
The momentum $p$ and energy $E$ of the bound state are
\beq
e^{ip}=\frac{x^+}{x^-}\,,
\qquad
E=Q+2ig\left(\frac{1}{x^+}-\frac{1}{x^-}\right)
=\sqrt{Q^2+16g^2\sin^2\frac{p}{2}}\,.
\eeq
It is useful to introduce the generalized rapidity $z$ defined on a torus with half periods
\beq
\omega_1=2\bK\,,
\qquad
\omega_2=2i\bK'-2\bK
\eeq
where $\bK$ and $\bK'$ are complete elliptic integral of the first kind with modulus squared 
$k^2=-16g^2/Q^2$ and $k'^2=1-k^2$. For real $g$ the first period is real and the second 
is purely imaginary. This torus covers the four copies of the $u$ plane. 
The spectral parameters, momentum and energy are expressed 
in terms of Jacobi elliptic functions (with modulus $k$) of the rapidity $z$
\beq
x^\pm(z)=\frac{Q}{4g}\left(\frac{\cn z}{\sn z}\pm i\right)(1+\dn z)\,,
\quad
p(z)=2\am z\,,
\qquad
\sin\frac{p}{2}=\sn z\,,
\quad
E(z)=Q\dn z\,.
\eeq

For real $z$ both $x^\pm>1$, the momentum is real and the energy positive. Shifting 
$z\to \bar z=z+\omega_2$ is a crossing transformation which replaces $x^\pm\to1/x^\pm$ and 
reverses the signs of both the energy and momentum.

The shift $z\to \tilde z=z+\omega_2/2$ gives the mirror theory, where $x^-<1<x^+$ 
and both the energy and momentum are purely imaginary. In the mirror theory their 
roles are reversed and one defines real mirror momentum and mirror energy as
\beq
\tilde p=iE=k'Q\sc1\tilde z\,,
\qquad
\tilde E=ip=2i\arcsin\left(\frac{Q}{4ig}\cd\tilde\zeta\right).
\eeq

\subsection{Reflections and open spin--chains}
\label{sec:reflect}

The Wilson loop breaks the $\psu(2,2|4)$ symmetry to $\osp(4^*|4)$ and the symmetry 
preserved by the ground state of the spin--chain $\psu(2|2)^2$ to the diagonal component 
$\psu(2|2)$ with the supercharges \eqn{diag}. To write down the spin--chain system 
preserving this symmetry, one can use the the method of images. This was done for the 
case of the open spin--chain associated to the D5-brane domain walls in \cite{CRY}.

\begin{figure}[t]
\begin{center}
\epsfig{file= 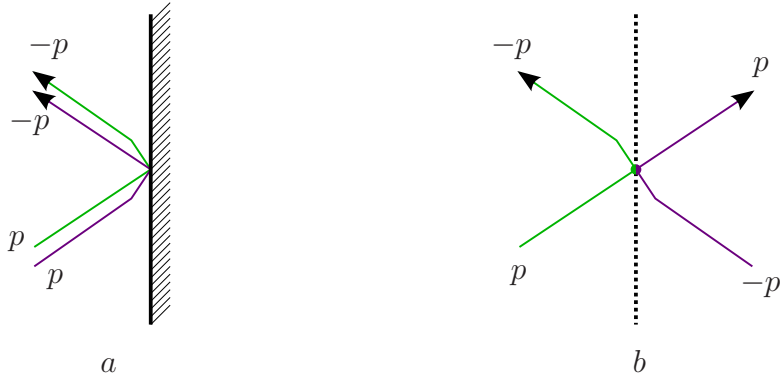,width=100mm
\psfrag{p}{$p$} 
\psfrag{-p}{$-p$} 
}
\\
$a\hskip2.7in b\ \ $\\
\parbox{15cm}{
\caption{Using the reflection trick a semi-infinite open spin--chain 
can be replaced with a spin--chain on the entire line. 
The original magnons ($a$) carry representations of 
$\psu(2|2)_R\times\psu(2|2)_L$ and momentum $p$, which gets 
reflected to momentum $-p$. In the doubled picture ($b$) the 
$\psu(2|2)_R$ label is carried by a magnon of momentum $-p$ on the right 
side, which gets scattered off the magnon of momentum $p$ and a 
$\psu(2|2)_L$ representation. As usual, the momentum does not get 
modified by the scattering and the magnon with momentum $-p$ continues 
on the left side, now carrying a $\psu(2|2)_L$ representation
\label{fig:doubling}}}
\end{center}
\end{figure}

The construction of the usual $\psu(2|2)$ scattering matrix is reviewed in Appendix~\ref{app:magnons}, 
following \cite{Beisert:2004ry}.

Consider first the right boundary and a semi-infinite spin--chain to its left. 
Each magnon with momentum $p$ carries a representation of 
$\psu(2|2)_L\times\psu(2|2)_R$. In the doubled description one
splits each of the magnons in two, one on the original left ray with momentum $p$ 
and a label of $\psu(2|2)_L$ and the image magnon 
of momentum $-p$ and a label of $\psu(2|2)_R$ on the right. 

Each magnon is charged under the three central charges of the extended 
$\psu(2|2)$: $C$, $P$ and $K$, related also to the labels $p$, $f$ and $a$ 
(see Appendix~\ref{app:magnons}). These charges are the same for the 
two copies of $\psu(2|2)$, and one should specify how they transform under 
the reflection. The reflection of the $\psu(2|2)_R$ magnons to the mirror 
description acts as
\bal
\label{S-reflect}
&p\to-p\,,
&\qquad
&x^\pm\to-x^\mp\,,
&\qquad
&a^2\to a^2\,,
&\qquad
&f\to fe^{ip}\,,
\\
&\bar P\to-\bar P\,,
&\qquad
&\bar K\to-\bar K\,,
&\qquad
&\bar Q\to i\bar Q\,,
&\qquad
&\bar S\to-i\bar S\,.
\eal
The extra phase acquired by $f$ is crucial, as for neighboring magnons the two $f$s 
have to satisfy $f_1=e^{ip_2}f_2$. 
For the magnons on the original (left) segment $f$ is given by the momenta 
of all the magnons to their right, {\em up to the wall}. For consistency, the image magnons 
on the right, should have $f$ which is the exponent of {\em minus} the momenta of all 
the magnons to their left up to the wall {\em including themselves}. The definition of 
$f$ above is exactly this, after the reflection.

In formulas, enumerating the magnons on the left of the wall $1,\cdots,M$, they satisfy 
$f_k=e^{ip_{k+1}}f_{k+1}$ with $f_M=1$. Then continue this pattern to the full real line, 
where the mirror magnons are labeled $M+1,\cdots,2M$. From \eqn{S-reflect} 
$f_{M+k+1}=e^{ip_{M-k}}f_{M-k}$ so that 
$f_{M+k}=e^{ip_{M-k+1}}f_{M-k+1}=f_{M-k}=e^{ip_{M+k+1}}f_{M+k+1}$ satisfies the 
correct relation.

If one considers the two particle scattering matrix, the order of the two entries is reversed as 
are the signs of the momenta. Instead of $f_2$ it will depend on the transformed 
$f_1\to e^{ip_1}f_1=e^{ip_1+ip_2}f_2$. Indeed it is easy to check that the S-matrix 
\eqn{S-matrix1}, \eqn{S-matrix2} is invariant 
under the transformation \eqn{S-reflect}
\beq
\bS(e^{ip_1+ip_2}f_2,-p_2,-p_1)=\bS(f_2,p_1,p_2)\,.
\eeq

The open $\psu(2|2)^2$ spin--chain on the left ray is the same as a $\psu(2|2)$ 
spin--chain on the full line. The symmetry generators in \eqn{diag} are the diagonal 
components of the original $\psu(2|2)_L$ and the reflection \eqn{S-reflect} of 
$\psu(2|2)_R$. The scattering of the right most magnon on the left part off its 
image, the left most magnon on the right part is then the reflection matrix for the 
open spin--chain. The matrix structure of this reflection matrix is fixed 
by extended $\psu(2|2)$ symmetry to be the same as the bulk reflection 
matrix
\beq
\bR^{(R)}{}^{a\dot a}_{b\dot b}(p)=R^{(R)}_0(p)\,\hat \bS^{\dot aa}_{\dot bb}(p,-p)\,.
\label{reflect}
\eeq
where
\beq
\hat \bS^{ab}_{cd}(p,-p)=\frac{\bS^{ab}_{cd}(p,-p)}{S_0(p,-p)}
\eeq
is the scattering matrix with the scalar factor removed. The explicit components 
of $\bR^{(R)}(p)$ are written down in Appendix~\ref{app:reflect}. 
The scalar part of the reflection matrix is not fixed by symmetry, but it is constrained 
by crossing symmetry, as discussed shortly.

The boundary Yang-Baxter equation is automatically satisfied, since it can be 
decomposed into several applications of the bulk Yang-Baxter equation for the 
$\psu(2|2)$ chain. It was also checked explicitly in \cite{CRY}.

\subsection{Boundary scalar factor}
\label{sec:dressing}

\begin{figure}[t]
\begin{center}
\epsfig{file= 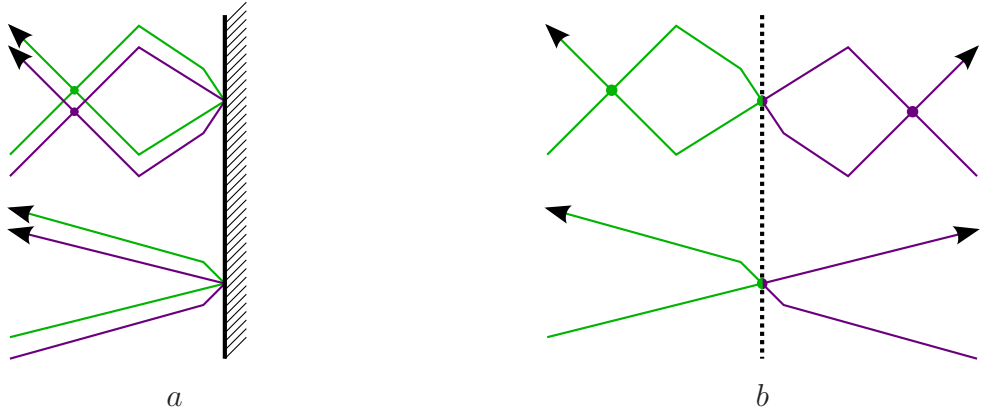,width=130mm
\psfrag{p}{$p$} 
\psfrag{-p}{$-p$} 
}
\\
$a\hskip3in b\ \ \ \ \ $\\
\parbox{15cm}{
\caption{The boundary crossing equation equates the two processes in ($a$). 
The bottom is a single reflection. The top has a bulk scattering (a pair of $\psu(2|2)$ 
scatterings, note that not four, since the two copies of $\psu(2|2)$ interact only at the 
boundary), and a crossed reflection. In the doubled picture ($b$) this single scattering 
(bottom) is related to a pair of regular scatterings and one with both particles crossed.
\label{fig:cross}}}
\end{center}
\end{figure}

Following \cite{Janik-cross} it is now commonly accepted that the scattering matrix of 
the bulk spin--chain is invariant under crossing symmetry. The same would then 
be expected also for the boundary reflection. In terms of generalized rapidity 
variables $z$, the boundary crossing unitarity condition states
\beq
\bR^{(R)}(z)=\bS(z,-z){\bR^{(R)}}^c(\omega_2-z)
\eeq
where ${\bR^{(R)}}^c$ is the reflection matrix in the crossed channel (see Figure~\ref{fig:cross}). 
For the open spin--chain model associated with giant gravitons this was solved in \cite{Ahn:2007bq}. 
In our case the bulk $S$ matrix is made of two $\psu(2|2)$ matrices and including 
all the indices this is
\beq
\label{boundarycross}
\bR^{(R)}{}_{a\dot a}^{b\dot b}(z)
=\bS_{ac}^{bd}(z,-z)\bS_{\dot a\dot c}^{\dot b\dot d}(z,-z)
\cC^{c\bar c}\cC_{d\bar d}\cC^{\dot c\dot{\bar c}}\cC_{\dot d\dot{\bar d}}
\bR^{(R)}{}_{\bar c\dot{\bar c}}^{\bar d\dot{\bar d}}(\omega_2-z)
\eeq
where $\cC^{c\bar c}$ is the charge conjugation matrices.

Deriving the equation for the boundary scalar phase is actually not that hard, 
using the crossing invariance of the bulk scattering 
matrix. The crossed reflection matrix is the same as the doubly crossed 
$\psu(2|2)$ scattering matrix. After including the usual scalar phase it is 
on its own invariant under both the crossings, giving
\bal
\label{explicit-ross}
&\cC^{c\bar c}\cC_{d\bar d}\cC^{\dot c\dot{\bar c}}\cC_{\dot d\dot{\bar d}}
\,\bR^{(R)}{}_{\bar c\dot{\bar c}}^{\bar d\dot{\bar d}}(\omega_2-z)
=
\frac{R_0(\omega_2-z)}{S_0(\omega_2-z,z-\omega_2)}
\cC^{c\bar c}\cC_{d\bar d}\cC^{\dot c\dot{\bar c}}\cC_{\dot d\dot{\bar d}}
\,\bS_{\bar c\dot{\bar c}}^{\bar d\dot{\bar d}}(\omega_2-z,z-\omega_2)
\\
&\qquad=
\frac{R_0(\omega_2-z)}{S_0(\omega_2-z,z-\omega_2)}
\cC^{\dot c\dot{\bar c}}\cC_{\dot d\dot{\bar d}}
\,\bS_{\dot{\bar c}d}^{\dot{\bar d}c}(z-\omega_2,-z)
=
\frac{R_0(\omega_2-z)}{S_0(\omega_2-z,z-\omega_2)}
\,\bS_{d\dot d}^{c\dot c}(-z,z)\,.
\eal
Expressing also the reflection matrix on the left hand side of \eqn{boundarycross} 
in terms of the bulk scattering matrix and using unitarity gives
\beq
\label{R0S0}
R_0(z)R_0(z-\omega_2)=S_0(z,-z)S_0(z-\omega_2,\omega_2-z)\,.
\eeq
This equation can be solved with $R_0(z)=S_0(z,-z)$.

This expression, though needs to be properly defined. The reason is that the 
solution of the crossing equation was formulated in terms of the dressing 
factor $\sigma(z_1,z_2)$, which is related to $S_0(z_1,z_2)$ by
\beq
S_0(z_1,z_2)^2=\frac{x_1^+-x_2^-}{x_1^--x_2^+}\frac{1-1/x_1^-x_2^+}{1-1/x_1^+x_2^-}
\,\frac{1}{\sigma(z_1,z_2)^2}
\eeq
and in particular
\beq
\label{sqrt}
S_0(z,-z)\sim\frac{x^+}{x^-}\sqrt{\frac{x^-+1/x^-}{x^++1/x^+}}
\,\frac{1}{\sigma(z,-z)}\,.
\eeq
So in defining $S_0(z,-z)$ one has to specify how to deal with the square root. 

It is natural therefore to take an ansatz for the boundary phase as
\beq
R_0(z)=\frac{\sigma_B(z)}{\sigma(z,-z)}
\eeq
such that equation \eqn{R0S0} is
\beq
\label{sigsig}
\sigma_B(z)\sigma_B(z-\omega_2)
=S_0(z,-z)S_0(z-\omega_2,\omega_2-z)
\sigma(z,-z)\sigma(z-\omega_2,\omega_2-z)\,.
\eeq
Using \eqn{sqrt} this last equation and the unitarity constraint become
\beq
\label{crossing}
\sigma_B(z)\sigma_B(z-\omega_2)= \frac{x^-+1/x^-}{x^++1/x^+}\,,
\qquad
\sigma_B(z)\sigma_B(- z)=1\,.
\eeq
This equation is solved in Appendix~\ref{app:dress}.

The above derivation was not very careful in treating the square--root branch cut in 
\eqn{sqrt}, so there may be a sign ambiguity on the right hand side of the first equation 
in \eqn{crossing}. The equation as 
written is in fact correct, to see that recall that the crossing equation was originally 
written for $S_0$ in \cite{Janik-cross} as
\bal
S_0(z_1+\omega_2,z_2)S_0(z_1,z_2)&={1}/{f(z_1,z_2)}
\\
S_0(z_1,z_2-\omega_2)S_0(z_1,z_2)&={1}/{f(z_1,z_2)}
\eal
and the dressing factor satisfies a very similar set of equations
\bal
\sigma(z_1+\omega_2,z_2)\sigma(z_1,z_2)&=\frac{x_2^-}{x_2^+}f(z_1,z_2)
\\
\sigma(z_1,z_2-\omega_2)\sigma(z_1,z_2)&=\frac{x_1^+}{x_1^-}f(z_1,z_2)
\eal
where in both cases
\beq
f(z_1,z_2)= \frac{x_1^--x_2^+}{x_1^--x_2^-}\frac{1-1/x_1^+x_2^+}{1-1/x_1^+x_2^-}
\eeq
Applying the monodromies twice on \eqn{sigsig} gives
\beq
\sigma_B(z)\sigma_B(z-\omega_2)
=\left(\frac{x^-}{x^+}\right)^2S_0(z,-z)^2\sigma_0(z,-z)^2
\eeq
so indeed $\sigma_B(z)$ should satisfy \eqn{crossing}

A solution found in Appendix~\ref{app:dress} following the methods of \cite{Volin:2009uv} is
\beq
\hat\sigma_B(z)=e^{i\chi_B(x^+)-i\chi_B(x^-)}
\eeq
where
\beq
\chi_B(x)=-i\oint\frac{dy}{2\pi i}\,
\frac{1}{x-y}\log \frac{\sinh 2\pi g(y+1/y)}{2\pi g(y+1/y)}\,.
\eeq
A more general solution to the crossing equation is gotten by multiplying by the exponent 
of an odd function of $p$, so $\hat\sigma_B e^{f_\text{odd}(p)}$ is also a solution. In particular 
the simplest modification is a linear function%
\footnote{Recall that $x^+/x^-=e^{ip}$, so if one uses $\hat\sigma_B$ as the definition of the 
square root in \eqn{sqrt}, then indeed $S_0(z,-z)=e^{ip}\hat\sigma_B(z)/\sigma(z,-z)$.}
$\hat\sigma_B e^{inp}$. In \cite{CMS} it was pointed out that also the replacement 
\beq
\hat\sigma_B\to \left(\frac{x^-+1/x^-}{x^++1/x^+}\right)^s \hat\sigma_B^{1-2s}
\eeq
still solves the equation \eqn{crossing}. 

To find the correct value of $s$ it is useful to compare with the reflection of magnons in 
string theory and compare it to the leading behavior of $\hat\sigma_B$ at strong coupling. 
The latter can be extracted from the first term in the expansion \eqn{ds}
\beq
\hat\sigma_B(z)\approx
\exp\left[
2ig\left(x^++\frac{1}{x^+}\right) \log\left(\frac{x^+-i}{x^++i}\right)
-2ig\left(x^-+\frac{1}{x^-}\right)\log \left(\frac{x^--i}{x^-+i}\right)\right]
\eeq
At leading order at strong coupling $x^\pm\approx e^{\pm ip/2}$, so
\beq
\hat\sigma_B(p)
\approx\exp\left[4ig\cos\frac{p}{2}\log\frac{1-\sin\frac{p}{2}}{1+\sin\frac{p}{2}}\right]
\eeq
Together with the usual strong coupling dressing factor restricted to the boundary 
$1/\sigma(p,-p)$ \cite{Arutyunov:2004vx,Hofman:2007xp}, the total reflection phase from the boundary at strong 
coupling is
\beq
R_0(p)
\approx
\exp\left[-8ig\cos\frac{p}{2}\log\cos\frac{p}{2}
+(1-2s)4ig\cos\frac{p}{2}\log\frac{1-\sin\frac{p}{2}}{1+\sin\frac{p}{2}}\right].
\eeq
The analogue calculations can be done in string theory by scattering in the 
classical solution of \cite{dk-spinchain} along the lines of \cite{Hofman:2007xp}. In addition 
to the usual sine-gordon scattering along the sphere part of the sigma model one 
needs to include a sinh-gordon contribution from the $AdS$ part.

Indeed the expressions agree for $s=1$. With an extra $e^{ip}=x^+/x^-$ the correct dressing phase is%
\footnote{Though this expression was not written down in the first version of this manuscript, 
the final expressions derived were correct, due to other sign errors.}
\beq
\label{sigmaB}
\sigma_B(z)=\frac{1+1/(x^-)^2}{1+1/(x^+)^2}\,e^{-i\chi_B(x^+)+i\chi_B(x^-)}
\eeq

The derivation of the boundary scattering phase 
above and in the appendix is for a fundamental magnon. For a bound state 
of $Q$ magnons it is defined as $R^Q_0(z)=\sigma^Q_B(z)/\sigma^{Q,Q}(z,-z)$. It is 
evaluated by considering the $Q(Q-1)$ scatterings of the constituent magnons of 
each-other and the $Q$ reflections of the constituents. $\sigma^{Q,Q}$ is then the 
product of $Q^2$ fundamental dressing factors and is the same as the usual bulk 
bound state dressing factor. $\sigma^Q_B$ is the product of $Q$ fundamental boundary 
dressing factors which ends up being identical to \eqn{sigmaB} with the appropriate 
$x^\pm$ for the bound state.

\subsection{Twisted boundary conditions}
\label{sec:twisted}

To construct the finite size open spin--chain one has to impose boundary conditions 
at the left end of the chain as well. The left boundary conditions are also associated 
to a Wilson loop, so are very similar to the right boundary conditions.%
\footnote{That is not required, of course, and it would be fun to consider open 
spin--chains with boundary conditions associated to different objects in the 
gauge theory, describing for example local insertions at the endpoint of 
an open Wilson loop, as in \cite{Aharony:2008an}.}
If the Wilson loop is straight, 
such that the angles $\theta=\phi=0$, the boundary conditions are completely 
compatible, and the left reflection matrix is identical to the right reflection matrix.%
\footnote{The phase $f$ also matches on this wall, since the total momentum 
including the magnons and their images is zero.}
The resulting open spin--chain is then described as a single $\psu(2|2)$ spin--chain 
with periodic boundary conditions and a symmetry requirement, that for every magnon 
of momentum $p$ on the original segment there is another magnon with momentum 
$-p$ on the mirror segment, as was also the case for the $SU(2)$ sector in 
\cite{dk-spinchain}.

When the angles $\theta$ and $\phi$ are different from zero, the two segments of Wilson 
loop to which the local operator is attached are not aligned. This means that the boundary 
conditions are not the same, but have a relative rotation with respect to each other. 
In the identification of $\psu(2|2)_L$ and $\psu(2|2)_R$ there would be different 
matrices instead of the $\sigma^3$ appearing in \eqn{diag}. This can be implemented on the 
reflection matrix \eqn{reflect} by a $U(1)^2\subset SU(2|2)$ rotation. For example choose 
the rotation to act on the fundamental representation of $\psu(2|2)_R$ by
\beq
\label{twist}
w_1\to e^{i\theta}w_1\,,
\qquad
w_2\to e^{-i\theta}w_2\,,
\qquad
\vartheta_1\to e^{i\phi}\vartheta_1\,,
\qquad
\vartheta_2\to e^{-i\phi}\vartheta_2\,.
\eeq
This is just the representation matrix of the spacial rotation by angle $\phi$ and $R$-rotation 
by angle $\theta$. This twist matrix is labeled $\bG$ such that the reflection matrix from the 
left boundary becomes
\beq
\bR^{(L)}{}^{a\dot a}_{b\dot b}(p)=\bG^{-1}{}^{\dot a}_{\dot c}\,\bS^{a\dot c}_{\dot db}(-p,p)\,\bG_{\dot b}^{\dot d}\,.
\label{left-reflect}
\eeq

This is be very similar to the twisting arising in the cases 
of $\beta$ and $\gamma$ deformed $\cN=4$ SYM \cite{Beisert:2005if}. These models were discussed 
recently from the point of view of the TBA and the $Y$-system in 
\cite{Ahn:2010yv,Gromov:2010dy,Arutyunov:2010gu, Ahn:2011xq,deLeeuw:2012hp}. 
The twist appears in the TBA equations as chemical potential terms.

\begin{figure}[t]
\begin{center}
\epsfig{file=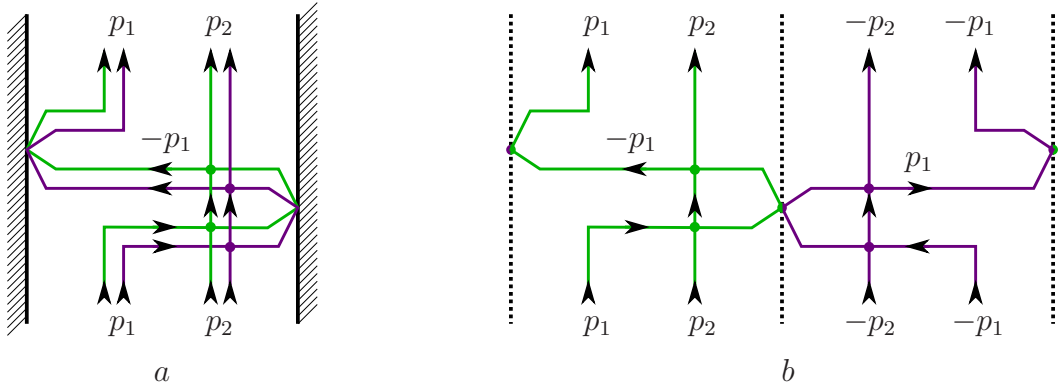,width=140mm
\psfrag{p1}{$p_1$}
\psfrag{p2}{$p_2$}
\psfrag{-p1}{$-p_1$}
\psfrag{-p2}{$-p_2$}
} 
$a\hskip3.2in b\hskip.6in$
\parbox{15cm}{
\caption{A graphical representation of the boundary Bethe-Yang equation in 
the open system with two boundaries ($a$), or in the doubled pictures ($b$) 
with periodic identification.
\label{fig:bbye}}}
\end{center}
\end{figure}

It is now possible to write the boundary Bethe-Yang equations for this system. 
It states that each magnon, if scattered with the others, reflected off the right boundary, 
scattered back all the way to the left boundary and then back to its original position 
(see Figure~\ref{fig:bbye}), it would pick up a trivial phase. 
In matrix form for a spin--chain with $L$ sites they are
\beq
e^{2i Lp_i}
=\prod_{j=i+1}^M\bS(p_i, p_j)\bR^{(R)}(p_i)
\prod_{\genfrac{}{}{0pt}{}{j=M}{j\neq i}}^1\bS(p_j, -p_i)\bR^{(L)}(-p_i)
\prod_{j=1}^{i-1}\bS(p_i, p_j)\,.
\eeq
As usual this set of equations can be diaganolized by using the nested Bethe ansatz equations, 
as was done for the case of the D5-brane defects in \cite{CRY}. The set of equations for the case 
at hand is almost identical, one just has to insert the extra phases from $\bG$, which acts 
diagonally on the nested equations.

In addition to the fundamental magnons the spin--chain has bound states. In the physical 
theory they are in totally symmetric representation of each of the $\psu(2|2)$ algebras. 
The $Q^\text{th}$ symmetric representation can 
be written in terms of homogeneous polynomials of degree $Q$ in the two bosonic and 
two fermionic variables $w_1$, $w_2$, $\vartheta_3$ and $\vartheta_4$. The representation is 
comprised of
\bal
&\text{Bosons:}\quad
&&w_1^lw_2^{Q-l}\,;\quad l=0,\cdots,Q\,,
\qquad
&&w_1^lw_2^{Q-2-l}\vartheta_3\vartheta_4\,;\quad l=0,\cdots,Q-2\,,
\\
&\text{Fermions:}\quad
&&w_1^lw_2^{Q-1-l}\vartheta_3\,;\quad l=0,\cdots,Q-1\,,
\qquad
&&w_1^lw_2^{Q-1-l}\vartheta_4\,;\quad l=0,\cdots,Q-1\,,
\eal
The states of the mirror model are of more importance in what follows. They are in the 
totally antisymmetric representations \cite{Arutyunov:2007tc}, which are homogenous 
polynomials in a pair of fermionic and a pair of bosonic variables with the opposite lables
$\vartheta_1$, $\vartheta_2$, $w_3$ and $w_4$. The states and their transformation 
under $\bG$ are
\bal
&\text{Bosons:}&&
\\
&&&w_3^lw_4^{Q-l}\to e^{i(2l-Q)\phi}\,w_3^lw_4^{Q-l}\,;
\qquad &&l=0,\cdots,Q\,,
\\
&&&w_3^lw_4^{Q-2-l}\vartheta_1\vartheta_2\to e^{i(2l+2-Q)\phi}\,w_3^lw_4^{Q-2-l}\vartheta_1\vartheta_2\,;
\qquad&& l=0,\cdots,Q-2\,,
\\
&\text{Fermions:}
\\
&&&w_3^lw_4^{Q-1-l}\vartheta_1\to e^{i(2l+1-Q)\phi+i\vartheta}\,w_3^lw_4^{Q-1-l}\vartheta_1\,;
\qquad &&l=0,\cdots,Q-1\,,
\\
&&&w_3^lw_4^{Q-1-l}\vartheta_2\to e^{i(2l+1-Q)\phi-i\vartheta}\,w_3^lw_4^{Q-1-l}\vartheta_2\,;
\qquad &&l=0,\cdots,Q-1\,,
\eal
Then it is easy to calculate the supertrace
\bal
\label{supertrace}
\sTr_Q\bG
&=\sum_{l=0}^Qe^{(2l-Q)i\phi}
+\sum_{l=0}^{Q-2}e^{(2l+2-Q)i\phi}
-2\cos\theta\sum_{l=0}^{Q-1}e^{(2l+1-Q)i\phi}
\\
&=2(\cos\phi-\cos\theta)\sum_{l=0}^{Q-1}e^{(2l+1-Q)i\phi}
=2(\cos\phi-\cos\theta)\frac{\sin Q\phi}{\sin\phi}
\eal

\section{Wrapping corrections}
\label{sec:wrapping}

\begin{figure}[t]
\begin{center}
\epsfig{file=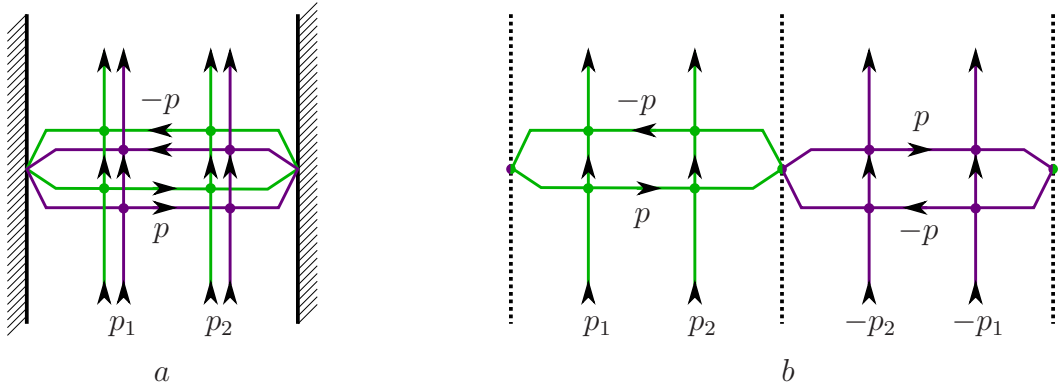,width=140mm
\psfrag{p}{$p$}
\psfrag{-p}{$-p$}
\psfrag{p1}{$p_1$}
\psfrag{p2}{$p_2$}
\psfrag{-p1}{$-p_1$}
\psfrag{-p2}{$-p_2$}
} 
$a\hskip3.2in b\hskip.6in$
\parbox{15cm}{
\caption{The double--row transfer matrix  
evaluated on a state with two physical magnons of momentum 
$p_1$ and $p_2$ ($a$). The auxiliary magnon, of momentum $p$ scatters 
off them, reflects off the boundary, scatters again and then does a crossed 
reflection back to the original starting point, where it can be traced. 
In the doubled version ($b$) there are a pair of auxiliary $\psu(2|2)$ magnons 
of momentum $p$ and $-p$. They scatter off all the physical magnons and 
their images as well as off each other. Once in the forward direction and once 
doubly crossed.
\label{fig:row}}}
\end{center}
\end{figure}

An elegant way to formulate the Bethe-Yang equations is in terms of the transfer 
matrix, capturing the monodromy around the spin--chain. In the case of open 
spin--chains the analog quantity is known as the double--row transfer matrix 
\cite{Cherednik:1985vs,Sklyanin:1988yz} defined (up to normalization) as
\beq
\bT(z|z_1,\cdots, z_M)\propto\sTr\Big[\bS(z,z_1)\cdots \bS(z,z_M)\bR^{(R)}(z)
\bS(z_M,-z)\cdots\bS(z_1,-z){\bR^{(L)}}^c(z-\omega_2)\Big]
\eeq
where the trace is performed only over the states of the magnon with generalized rapidity 
$z$ and it carries the matrix indices of all the other magnons.%
This is illustrated in Figure~\ref{fig:row}.

\begin{figure}[t]
\begin{center}
\epsfig{file=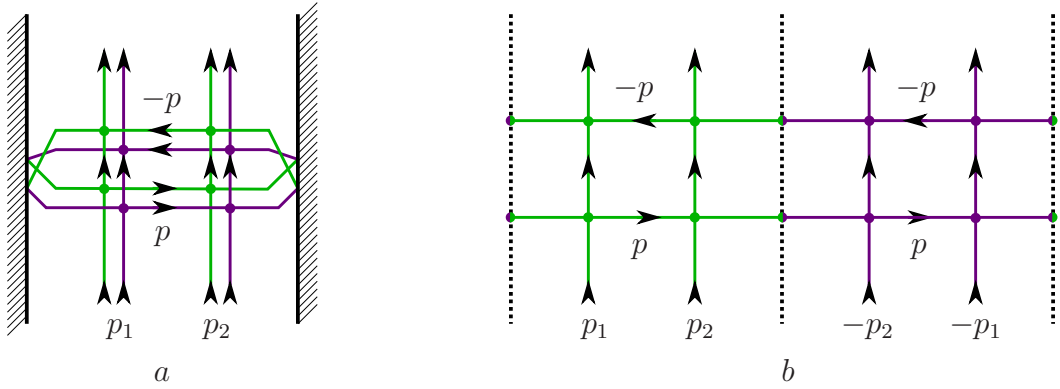,width=140mm
\psfrag{p}{$p$}
\psfrag{-p}{$-p$}
\psfrag{p1}{$p_1$}
\psfrag{p2}{$p_2$}
\psfrag{-p1}{$-p_1$}
\psfrag{-p2}{$-p_2$}
} 
$a\hskip3.2in b\hskip.6in$
\parbox{15cm}{
\caption{An alternative description of the double row transfer matrix, 
see the sequence of steps in Figure~\ref{fig:ybe}. 
The result in the doubled picture ($b$) is a pair of regular $\psu(2|2)$ transfer matrices.
The only remnant of the boundary reflections is the phase factor 
$\sigma_B$ at each boundary and possibly a twist matrix. In 
the open spin--chain picture ($a$) the boundary reflection is completely diagonal.
\label{fig:row1}}}
\end{center}
\end{figure}

\begin{figure}[t]
\begin{center}
\epsfig{file=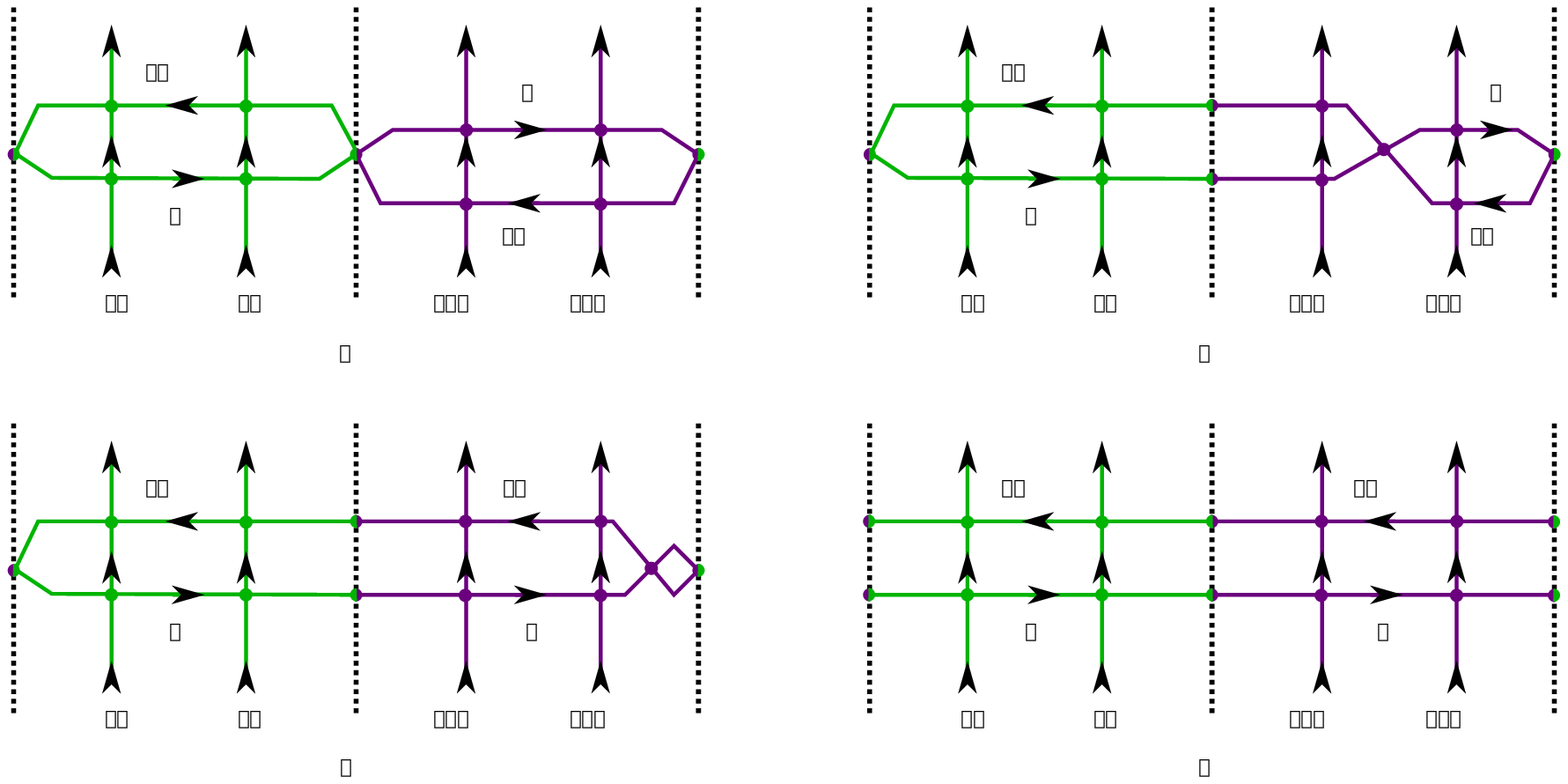,width=140mm
\psfrag{p}{$p$}
\psfrag{-p}{$-p$}
\psfrag{p1}{$p_1$}
\psfrag{p2}{$p_2$}
\psfrag{-p1}{$-p_1$}
\psfrag{-p2}{$-p_2$}
\psfrag{a}{$a$}
\psfrag{b}{$b$}
\psfrag{c}{$c$}
\psfrag{d}{$d$}
} 
\parbox{15cm}{
\caption{The sequence of Yang-Baxter and crossing/unitarity relations 
$a\to b\to c\to d$ used to relate Figure~\ref{fig:row} and Figure~\ref{fig:row1}. 
Recall that up to a scalar phase and a twist with which it commutes, 
the reflection matrix is the same as the regular $\psu(2|2)$ scattering matrix.
\label{fig:ybe}}}
\end{center}
\end{figure}

In terms of the transfer matrix the Bethe-Yang equations take the very simple form 
$\bT(z_i|z_1,\cdots,z_M)e^{-2iLp_i}=-1$.

Within the context of integrability of $\cN=4$ SYM the double row transfer matrix 
was studied in \cite{Murgan:2008fs, Galleas:2009ye} 
and used in calculating finite size corrections to 
the spectrum of insertions into determinant operators (giant gravitons in the 
dual string theory) in \cite{Correa:2009mz,Bajnok:2010ui}, generalizing 
the pioneering work of Bajnok-Janik \cite{bajnok-janik}. Here the 
same is done for the open spin--chain model related to Wilson loops.

It is easy to show that in the case studied here, where the reflection matrix is proportional 
to the bulk scattering matrix and commutes with the twist matrix, 
the transfer matrix factorizes to the product of 
two twisted transfer matrices of $\psu(2|2)$ as in Figure~\ref{fig:row1}. 
Instead of writing formulas with intractable 
indices, the reader should be convinced by Figure~\ref{fig:ybe}. 
The twisted transfer matrices of the closed spin--chain were studied in 
\cite{Arutyunov:2010gu}, see also 
\cite{Ahn:2010yv,Gromov:2010dy,Ahn:2010ws,deLeeuw:2012hp}.

The main example of interest is the ground state energy, 
so we can omit all the magnons except for the auxiliary one being traced 
over. Using $\bR^{(L)}(z)=\bG^{-1}\,\bR^{(R)}(-z)\,\bG$ one finds
\bal
\bT(z)&\propto\sTr\Big[\bR^{(R)}(z)\,{\bR^{(L)}}^c(z-\omega_2)\Big]
=\sTr\Big[\bR^{(R)}(z)\,\bG^{-1}\,{\bR^{(R)}}^c(-z+\omega_2)\,\bG\,\Big]
\\&
=\frac{R_0(z)R_0(-z+\omega_2)}
{S_0(z,-z)S_0(-z+\omega_2,z-\omega_2)}
\,\bS_{a\dot a}^{\dot bb}(z,-z)\,\bG^{-1}{}_{\dot b}^{\dot c}
\,\bS_{\dot c b}^{a\dot d}(-z,z)\,\bG_{\dot d}^{\dot a}
\\&
=\sigma_B(z)\sigma_B(-z+\omega_2)\left(\frac{x^-}{x^+}\right)^2(\sTr \bG)^2
\eal
where the same manipulations as in \eqn{explicit-ross} and \eqn{sigsig} were employed.
The final expression is very simple because the twist matrices $\bG$ commute 
with the scattering matrix and thus lead to the factorization into two traces of 
the twist matrix and a scalar phase.

For the purpose of calculating the wrapping corrections one needs to properly normalize this 
as in \eqn{second-T}. This end up giving rougly the inverse kinematic factor, so the 
transfer function $T_Q^{\phi,\theta}(z)$ for an auxiliary bound state 
of $Q$ magnons with rapidity $z$ and with the twist angles explicitly spelled out is
\bal
\label{kappa}
T_Q^{\phi,\theta}(z)
&=\frac{1}{\sigma_B(z)\sigma_B(-z+\omega_2)}(\sTr \bG)^2
=\frac{\sigma_B(z-\omega_2)}{\sigma_B(z)}(\sTr \bG)^2
\\&
=\left(\frac{x^-}{x^+}\right)^2
e^{-i\chi_B(1/x^+(z))+i\chi_B(1/x^-(z))+i\chi_B(x^+(z))-i\chi_B(x^-(z))}
(\sTr \bG)^2
\eal
which used that $x^\pm(z-\omega_2)=1/x^\pm(z)$.

The auxiliary particle can be in any representation. For the purpose of calculating 
the wrapping effects it should be a bound state in the mirror model, which are 
in the completely antisymmetric representation. 
Each of the supertraces is performed in the 
$Q^\text{th}$ antisymmetric representation of $\psu(2|2)$ which using 
\eqn{supertrace} leads to
\beq
\left(\sTr_Q\bG\right)^2
=4(\cos\phi-\cos\theta)^2\frac{\sin^2Q\phi}{\sin^2\phi}
\eeq

The ground state $Z^J$ has classical dimension $\Delta=J$. The first correction 
comes from the leading wrapping effect, which is
\beq
\label{energy}
\delta E\approx-\frac{1}{2\pi}\sum_{Q=1}^\infty\int_0^\infty d\tilde p
\log\left(1+T^{(\phi,\theta)}_Q(z+{\textstyle\frac{\omega_2}{2}}) e^{-2L\tilde E_Q}\right)
\eeq
Here $\tilde z=z+\omega_2/2$ is the generalized rapidity for a magnon of the mirror 
theory of momentum $\tilde p$. $L$ is the length of the world-sheet and is equal 
to the number of sites $J$ on the spin--chain. Finally $\tilde E_Q=\log(x^+/x^-)$ is 
the energy of the bound state in the mirror theory. A similar expression arises 
from the full thermodynamic Bethe ansatz treatment of the problem, where 
$\tilde E_Q$ is replaced by $\tilde\epsilon_Q$, the solution of the TBA equations, 
accounting for entropy, in addition to the energy. In that case $L$ is interpreted 
as the inverse temperature. 

\newpage
The $(x^-/x^+)^2$ term in the scalar factor \eqn{kappa} can be absorbed in a shift 
$L\to L+1$ and for the other terms we note that in the mirror kinematics 
$x^+(z+\frac{\omega_2}{2})>1$ and $x^-(z+\frac{\omega_2}{2})<1$, so 
one should use the monodromy property \eqn{discontinuity} to write the expression 
in terms of $x^+$ and $1/x^-$. This gives
\bal
\label{sig-ratio}
&e^{2i\chi_B(x^+(z+\frac{\omega_2}{2}))+2i\chi_B(1/x^-(z+\frac{\omega_2}{2}))-2i\Phi_0}\,
\frac{(2\pi g)^2(x^++1/x^+)(x^-+1/x^-)}
{\sinh (2\pi g(x^++1/x^+))\sinh (2\pi g(x^-+1/x^-))}
\\&\qquad
=e^{2i\chi_B(x^+(z+\frac{\omega_2}{2}))+2i\chi_B(1/x^-(z+\frac{\omega_2}{2}))-2i\Phi_0}\,
\frac{(2\pi)^2(u^2+Q^2/4)}{\sinh^2(2\pi u)}
\eal
Where the last equality was written using the relation $g(x^\pm+1/x^\pm)=u\pm iQ/2$ and 
the periodicity of the hyperbolic sine function. Note that the dressing phase contribution has 
a pole at $u=0$, which is also $\tilde p=0$, representing the contribution of zero momentum 
states for all values of $Q$.

Usually one expands the log function in \eqn{energy} to linear order and integrates over $\tilde p$. 
In the case at hand this approximation is invalid, since $T_Q^{\phi,\theta}(\tilde p)$ 
has the double pole at $\tilde p=0$, 
and thus is not uniformly small for all $\tilde p$ (even for small $g$ or for large $L$). Instead 
write \cite{Bajnok:2004tq,Bajnok:2006dn}
\beq
T_Q=\frac{T_Q^\text{pole}}{\tilde p^2}+T_Q^\text{reg}\,,
\eeq
and use
\beq
\label{int-id}
\int_0^\infty d\tilde p\,\log\left(1+\frac{c^2}{\tilde p^2}\right)=\pi c\,,
\eeq
to get
\bal
\label{expand-log}
\delta E&=-\frac{1}{2\pi}\sum_{Q=1}^\infty\int_0^\infty d\tilde p
\log\left(1+\frac{T_Q^\text{pole}}{\tilde p^2} e^{-2L\tilde E_Q}\right)
-\frac{1}{2\pi}\sum_{Q=1}^\infty\int_0^\infty d\tilde p
\log\left(\frac{1+T_Q\, e^{-2L\tilde E_Q}}
{1+T_Q^\text{pole} e^{-2L\tilde E_Q}/\tilde p^2}
\right)
\\
&=-\frac{1}{2}\sum_{Q=1}^\infty\sqrt{T_Q^\text{pole} e^{-2L\tilde E_Q}}
-\frac{1}{2\pi}\sum_{Q=1}^\infty\int_0^\infty d\tilde p\,
\frac{T_Q^\text{reg} e^{-2L\tilde E_Q}}
{1+T_Q^\text{pole} e^{-2L\tilde E_Q}/\tilde p^2}
+\cdots
\eal
The pole term dominates at large $L$ and at weak coupling, since it comes with 
$e^{-L\tilde E_Q}$ rather than $e^{-2L\tilde E_Q}$. And it is given by the simple expression
\beq
\label{sqrt-T}
\sqrt{T_Q^\text{pole} e^{-2L\tilde E_Q}}
=
2\frac{\cos\phi-\cos\theta}{\sin\phi}\,\sin Q\phi\,
\mathop{\text{res}}_{\tilde p\to0}
\left[e^{i(\chi_B(x^+)+\chi_B(1/x^-))}
\,\frac{2\pi\sqrt{u^2+Q^2/4}}{(-1)^Q\sinh(2\pi u)}
\left(\frac{x^-}{x^+}\right)^{L+1}
\right]
\eeq
One may be concerned about the choice of sign on the right hand side. 
If the integral in \eqn{int-id} is regarded as a real integral of a positive definite 
quantity, then one should choose the positive branch of 
$\sqrt{T_Q^\text{pole} e^{-2L\tilde E_Q}}$. 
It is actually more natural to take it to be an analytic expression and 
then the hyperbolic sine in the denominator is $\sinh(2\pi (u\pm iQ/2))$. 
See a careful discussion of such signs in \cite{Bajnok:2006dn}.

By using the explicit expressions at small $\tilde p$
\bal
u&=\frac{\tilde p}{2}\sqrt{1+\frac{16 g^2}{\tilde p^2+Q^2}}
=\frac{\tilde p}{2}\sqrt{1+\frac{16 g^2}{Q^2}}+O(\tilde p^3)
\\
e^{-\tilde E_Q}&=\frac{x^-}{x^+}
=\frac{\sqrt{1+\frac{16 g^2}{\tilde p^2+Q^2}}-1}{\sqrt{1+\frac{16 g^2}{\tilde p^2+Q^2}}+1}
=\frac{\sqrt{1+\frac{16 g^2}{Q^2}}-1}{\sqrt{1+\frac{16 g^2}{Q^2}}+1}+O(\tilde p^2)
\eal
equation \eqn{sqrt-T} becomes
\beq
\sqrt{T_Q^\text{pole} e^{-2L\tilde E_Q}}
=
2\frac{\cos\phi-\cos\theta}{\sin\phi}\,\sin Q\phi\,
\frac{(-1)^QQ}{\sqrt{1+\frac{16 g^2}{Q^2}}}
\left(\frac{\sqrt{1+\frac{16 g^2}{Q^2}}-1}{\sqrt{1+\frac{16 g^2}{Q^2}}+1}\right)^{L+1}
e^{i(\chi_B(x^+)+\chi_B(x^-))}
\eeq
At weak coupling the dressing phase starts contributing at order $g^4$, so it can 
be ignored when considering the first two terms
\beq
\sqrt{T_Q^\text{pole} e^{-2L\tilde E_Q}}
=2\frac{\cos\phi-\cos\theta}{\sin\phi}\,\sin Q\phi\,(-1)^Q\left[
\frac{(4g^2)^{L+1}}{Q^{2L+1}}
-2(L+2)\frac{(4g^2)^{L+2}}{Q^{2L+3}}
+O(g^{2(L+3)})\right]
\eeq
And then from \eqn{expand-log}, the leading contribution to the ground state energy at small $g$ is
\bal
\delta E&\approx-(4g^2)^{L+1}\frac{\cos\phi-\cos\theta}{\sin\phi}
\sum_{Q=1}^\infty\frac{(-1)^Q\sin Q\phi}{Q^{2L+1}}
\\&
=-\frac{(4g^2)^{L+1}}{2i}\frac{\cos\phi-\cos\theta}{\sin\phi}
\left(\Li_{2L+1}(-e^{i\phi})-\Li_{2L+1}(-e^{-i\phi})\right)
\\&
=-\frac{(-16\pi^2g^2)^{L+1}}{4\pi(2L+1)!}\frac{\cos\phi-\cos\theta}{\sin\phi}
\,B_{2L+1}\left(\frac{\phi}{2\pi}+\frac{1}{2}\right)
\eal
Where $B_{2L+1}$ are Bernoulli polynomials. For $L=0$ it is $B_1(x)=x-\frac{1}{2}$, so
\beq
E=\delta E=2g^2\,\frac{\cos\phi-\cos\theta}{\sin\phi}\,\phi+O(g^4)
\eeq
This is the same as the one loop perturbative calculation, equation \eqn{pert}, as 
calculated originally in \cite{DGO}.

For $L>0$ this calculates the leading correction to the energy of the ground state $Z^L$ 
in the spin--chain, which happens at order $L+1$ in perturbation theory and should 
hold up to order $2L+2$, where double wrapping and the second 
term in \eqn{expand-log} start to contribute. For $L>0$ this 
would involve gauge theory graphs like in Figure~\ref{fig:graphs}$d$, which so 
far have not been calculated directly, though the final expression appeared already 
in \cite{CHMS2} and the perturbative calculation should be closely related.

For large $L$ (scaled with $g$) this should give the correct answer to 
all values of the coupling and match with classical string solutions similar to those in 
\cite{dk-spinchain}.

\section{The twisted boundary TBA}
\label{sec:BTBA}

The calculation so far enabled the derivation of the one--loop result for the generalized quark-antiquark 
potential. To find the answer at all values of the coupling requires to solve exactly for the 
ground state energy of the open spin--chain with twisted boundary conditions. This can be 
done by using the boundary thermodynamic Bethe ansatz (BTBA) equations 
\cite{LeClair:1995uf}.

The idea is to exchange the space and time directions and 
instead of calculating the partition function (or Witten index) 
of the finite size system over an infinite time, to calculate the partition 
function of the theory on an infinite circle over a finite time \cite{Z-TBA}. 
In the case of a periodic model the result is a thermal partition function
\beq
Z=\Tr e^{-R\,H_L}=\Tr e^{-H^m_R/T}\,,
\qquad
T=1/L\,,
\eeq
where $H_L$ is the hamiltonian for the original model on the interval of width $L$ and 
$H^m_R$ is the hamiltonian of the mirror model on the interval of width $R$.

\begin{figure}[t]
\begin{center}
\epsfig{file=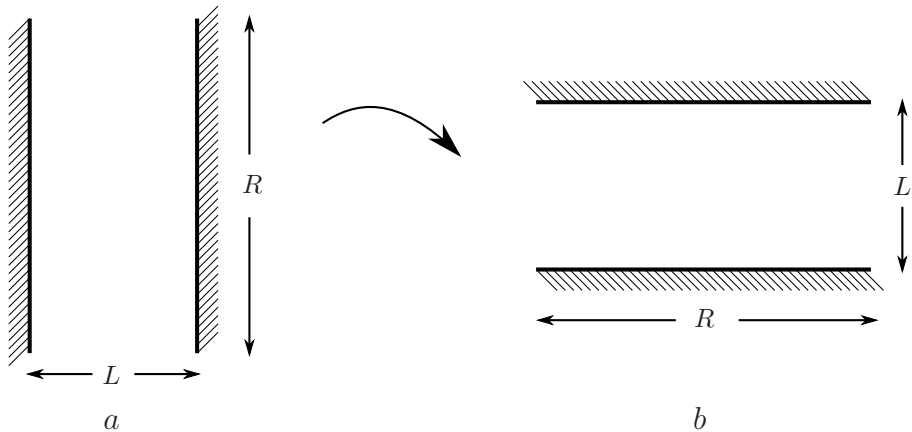,width=120mm
\psfrag{L}{\footnotesize$L$} 
\psfrag{R}{\footnotesize$R$} 
}
\\
$a\hskip3in b\hskip.55in $\\
\parbox{15cm}{
\caption{The boundary TBA: In the original setup ($a$) one has to find the 
exact ground state spectrum for finite size $L$. Instead, in the BTBA approach 
($b$) one considers the system on a very large cylinder of circumference $R$, 
where the spectrum can be evaluated exactly and propagates a boundary state 
over a finite time $L$.
\label{fig:btba}}}
\end{center}
\end{figure}

In the case of the open spin--chain the boundaries get replaced with initial and final 
boundary states $\ket{\cB_\alpha}$ and one calculates the transition amplitude 
between boundaries of type $\alpha,\beta$
\beq
Z_{\alpha,\beta}=\Tr e^{-R\,H_{L(\alpha,\beta)}}
=\bra{\cB_\beta}e^{-H^m_R/T}\ket{\cB_\alpha}\,,
\qquad
T=1/L\,,
\eeq
see Figure~\ref{fig:btba}.  
In general there are many different boundary conditions for the open spin--chain related 
to strings ending on different objects in $AdS_5\times\bS^5$. 
The ones addressed here are those related to insertions into Wilson loop operators 
as discussed in the previous sections and the two boundaries are identical up to 
the rotation by the angles $\phi$ and $\theta$.

The boundary state $\ket\cB$ is a superposition of states with different numbers of magnons 
of arbitrary momenta, subject to the symmetry under $\tilde p\to-\tilde p$. 
The amplitude for emitting a pair of magnons by the boundary is represented by 
the matrix $K(\tilde p)$ which is closely related to the reflection matrix $R$. 
It is the charge conjugate of the inverse reflection matrix analytically continued 
to the mirror kinematics%
\footnote{In this section quantities are written in terms of real mirror rapidity $\tilde z$.}
\beq
K^{a\dot a b\dot b}(\tilde z)
=\cC^{ac}\cC^{\dot a\dot c}R^{-1}{}_{c\dot c}^{b\dot b}(-\tilde z+\textstyle\frac{\omega_2}{2})\,,
\eeq
where $\cC$ is the charge conjugation matrix.

In addition, the boundary state can have single magnon contributions subject to the 
constraint $\tilde p=0$. These arise when the scattering matrix has a double pole at 
$\tilde p=0$, which is indeed the case here. The amplitude for the emission of such 
a magnon is $g^{a\dot a}$, related to $K$ by \cite{Dorey:2000eh}
\beq
\mathop{\Res}_{\tilde p\to0}K^{a\dot ab\dot b}(\tilde p)
=2i\,g^{a\dot a}g^{b\dot b}
\eeq
With these two quantities, and under the usual assumptions of integrability, 
the boundary state can be written as \cite{Gho-Z}
\beq
\label{boundary}
\ket{\cB}
=\cN\left(1+g^{a\dot a}A^\dagger_{a\dot a}(0)\right)
\exp\left[\int_0^{\omega_1/2} d\tilde z\, K^{a\dot ab\dot b}(\tilde z)
A^\dagger_{a\dot a}(-\tilde z)A^\dagger_{b\dot b}(\tilde z)\right]\ket{0}
\eeq
where $A^\dagger_{a\dot a}(\tilde p)$ is a creation operator for a magnon with quantum numbers 
$a$ and $\dot a$ in $\psu(2|2)_L\times\psu(2|2)_R$ and $\cN$ is a normalization constant. 
This expression was written in terms of ``out states'', where the momentum 
of the right magnon is larger than the left magnon. In terms of ``in states'' the 
same expression can be written by virtue of the boundary crossing relation 
\eqn{boundarycross} as
\beq
\ket{\cB}
=\cN\left(1+g^{a\dot a}A^\dagger_{a\dot a}(0)\right)
\exp\left[\int_0^{\omega_1/2} d\tilde z\, K^{a\dot ab\dot b}(-\tilde z)
A^\dagger_{a\dot a}(\tilde z)A^\dagger_{b\dot b}(-\tilde z)\right]\ket{0}
\eeq
Note that the in addition to fundamental magnons these expressions also includes bound states 
of the mirror model. If one wishes to be more explicit about that, it is possible to add an 
index $Q$ to $g$, $K$ and $A$ and then sum over $Q$.

The difference between the initial state and the final state, which can be labeled 
$\ket{\cB^{(R)}}$ and $\ket{\cB^{(L)}}$ is the same as the difference between the 
respective scattering matrices --- it is a global rotation.

Indeed, as mentioned before, when the two boundaries are of the type associated to the Wilson 
loops, then by repeated use of the Yang-Baxter equation and unitarity, it is possible to 
replace in all calculations the reflection matrix by a diagonal matrix with 
only the scalar phase and the twist. Then one gets
\bal
K^{(R)\,a\dot a b\dot b}(\tilde z)
=\frac{\cC^{ac}\delta^{\dot b}_c\,\cC^{\dot a\dot c}\delta^b_{\dot c}}{\sigma_B(-\tilde z+\textstyle\frac{\omega_2}{2})}
\qquad
\bar K^{(L)}{}_{a\dot a b\dot b}(\tilde z)
=\frac{\cC_{ac}\bG^{-1}{}_{\dot b}^c\,\cC_{\dot a\dot c}\bG_b^{\dot c}}{\sigma_B(-\tilde z+\textstyle\frac{\omega_2}{2})}
\eal
where $\bar K$ is the conjugate of $K$, the amplitude of absorption of a pair of 
magnons into the final state. Since the order of the operators is reversed, 
in the ``in states'' basis the amplitude is $\bar K(\tilde z)$ and in the ``out states'' basis 
it is $\bar K(-\tilde z)$.

The vacuum transfer matrix $T^{\phi,\theta}_Q$ is then the inner product of this two-particle contribution
\bal
\label{second-T}
T^{\phi,\theta}_Q
=\bar K^{(L)}_{a\dot a b\dot b}(-\tilde z)K^{(R)\,b\dot ba\dot a}(\tilde z)
&=\frac{\cC_{ad}\bG^{-1}{}^d_{\dot b}\,\cC_{\dot a\dot d}\bG^{\dot d}_b
\,\cC^{ac}\delta^{\dot b}_c\,\cC^{\dot a\dot c}\delta^b_{\dot c}}
{\sigma_B(\tilde z+\textstyle\frac{\omega_2}{2})\sigma_B(-\tilde z+\textstyle\frac{\omega_2}{2})}
\\&
=\frac{(\sTr\bG)^2}
{\sigma_B(\tilde z+\textstyle\frac{\omega_2}{2})\sigma_B(-\tilde z+\textstyle\frac{\omega_2}{2})}
=(\sTr\bG)^2\lambda_Q(z)
\eal
where after absorbing the factor of $(x^-/x^+)^2$ in \eqn{kappa} by the shift $L\to L+1$, 
$\lambda_Q$ is given by \eqn{sig-ratio}%
\beq
\lambda_Q=e^{2i(\chi_B(x^+)+\chi_B(1/x^-)-\Phi_0)}
\frac{(2\pi)^2(u^2+Q^2/4)}{\sinh^2(2\pi u)}
\eeq

A crucial ingredient in deriving the result in Section~\ref{sec:wrapping} is the fact that 
the scattering matrix has a pole at zero mirror momentum, seen here in $\lambda_Q$. 
This represents 
contributions from single zero momentum states and the factor $g^{a\dot a}$ in 
equation \eqn{boundary} is related to the residue. 
For a generic bound state of $Q$ magnons it is
\beq
g_Q^{a\dot a}
=\cC^{a\dot a}(-1)^Q\,\frac{Q}{\sqrt{1+\frac{16 g^2}{Q^2}}}
\frac{\sqrt{1+\frac{16 g^2}{Q^2}}-1}{\sqrt{1+\frac{16 g^2}{Q^2}}+1}
e^{i\chi_B(x^+)+i\chi_B(1/x^-)-i\Phi_0}
\eeq

The derivation of the boundary TBA equations follows the standard procedure 
\cite{LeClair:1995uf} of expressing 
the boundary state as a sum over eigenstate of the mirror hamiltonian, introducing 
densities for particles and holes and minimizing the free energy, accounting for the 
entropy of the states. As explained, the boundary state is completely symmetric 
in the exchange of $\tilde p\to-\tilde p$ and under the replacement of 
$\psu(2|2)_R$ and $\psu(2|2)_L$. The density of momentum carrying particles 
can therefore be defined for positive $\tilde p$ only, representing a pair of particles 
(or holes) at momentum $\tilde p$ and $-\tilde p$. At the nested level the 
replacement of $u\to-u$ is accompanied by replacement of the right and left 
groups.

The mirror theory is exactly the same as in the case of periodic spin--chains and the Bethe equations 
(which lead to the kernels in the TBA equations) are also the same, except that 
one has to account for the symmetry of the states overlapping with the boundary 
state.

The resulting BTBA equations are therefore the same as for the closed spin--chains 
\cite{Bombardelli:2009ns,Gromov:2009bc,AF-TBA} with the following modifications:
\begin{enumerate}
\item
The momentum carrying densities $\epsilon_Q$ are defined only for positive 
$u$.
\item
Since now the $\epsilon_Q$ densities represent pairs of particles of opposite momentum, 
all kernels coupling to them are doubled
$K^{QQ'}(u,u')\to K^{QQ'}(u,u')+K^{QQ'}(u,-u')$. The integration domain for $u'$ is 
$[0,\infty)$.
\item
All other densities are symmetric under
\beq
\epsilon^{(+)}_{y^\pm}(u)=\epsilon^{(-)}_{y^\pm}(-u)\,,
\qquad
\epsilon^{(+)}_{vw|M}(u)=\epsilon^{(-)}_{vw|M}(-u)\,,
\qquad
\epsilon^{(+)}_{w|M}(u)=\epsilon^{(-)}_{w|M}(-u)\,.
\eeq
\item
The twisting matrix $\bG$ introduces chemical potentials, as in the case of the usual 
close spin--chain twisted TBA. See \cite{Ahn:2010yv,Gromov:2010dy,Arutyunov:2010gu}.
\item
There is an extra driving term from the boundary dressing phase contributing 
to the equation for $\epsilon_Q$. This is equal to $-\log\lambda_Q$.
\item
The ground state energy is
\beq
\delta E=-\frac{1}{2\pi}\sum_{Q=1}^\infty\int_0^\infty du\,\frac{d\tilde p}{du}
\log\left(1+\lambda_Q\, e^{-\epsilon_Q}\right)
\eeq
\end{enumerate}

An alternative formulation extends the definition of $\epsilon_Q$ to negative $u$ by 
$\epsilon_Q(-u)=\epsilon_Q(u)$ and then integrate $u$ over $(-\infty,\infty)$ 
with the regular kernels. In this formulation the BTBA equations are identical to the usual 
twisted periodic TBA equations except for the extra driving term $-\log\lambda_Q$ for 
$Y_Q$. In the notations of \cite{AF-TBA} the equations 
are
\begin{align}
\log Y^{(a)}_{w|M}&=-2i\theta+\log\left(1+1/Y^{(a)}_{w|N}\right)\star K_{NM}
+\log\frac{1-1/Y^{(a)}_-}{1-1/Y^{(a)}_+}\;\widehat{\star}\; K_{M}
\nonumber\\
\log Y^{(a)}_{vw|M}&=-2i\phi+\log\left(1+1/Y^{(a)}_{vw|N}\right)\star K_{NM}
+\log\frac{1-1/Y^{(a)}_-}{1-1/Y^{(a)}_+}\;\widehat{\star}\; K_{M}
-\log\left(1+Y_Q\right)\star K^{QM}_{xy}
\nonumber\\
\log Y^{(a)}_\pm&=-i(\phi-\theta)-\log\left(1+Y_Q\right)\star K^{QM}_\pm
+\log\frac{1+1/Y^{(a)}_{vw|M}}{1+1/Y^{(a)}_{w|M}}\star K_{M}
\\
\log Y_Q&=-2iQ\phi-2(L+1)\tilde E_Q+\log \lambda_Q
+\log\left(1+Y_{Q'}\right)\star K^{Q'Q}_{sl(2)}
\nonumber\\
&\quad
+\sum_{a=\pm}
\left[\log\left(1+1/Y^{(a)}_{vw|M}\right) \star K^{MQ}_{vwx}
+\sum_\pm\log\left(1-1/Y^{(a)}_\pm\right)\;\widehat{\star}\; K^{yQ}_\pm\right]
\nonumber
\end{align}
with
\beq
Y_Q=e^{-\epsilon_Q}\,,\qquad
Y^{(a)}_{vw|M}=e^{\epsilon^{(a)}_{vw|M}}\,,\qquad
Y^{(a)}_{w|M}=e^{\epsilon^{(a)}_{w|M}}\,,\qquad
Y^{(a)}_{y^\pm}=e^{\epsilon^{(a)}_{y^\pm}}\,.
\eeq
The definitions of the convolutions and all the kernels are given in \cite{AF-TBA} 
(though note the slightly different conventions from this manuscript).

The energy is then given by
\beq
E=-\frac{1}{4\pi}\sum_{Q=1}^\infty\int_{-\infty}^\infty du\, \frac{d\tilde p}{du}\log(1+Y_Q)\,.
\eeq

One can derive the simplified (and hybrid) TBA equations as usual (see \cite{Arutyunov:2009ax} 
by acting with the inverse kernel $(K+1)^{-1}_{NM}$ (see \cite{AF-TBA}). This is known to make 
the simplified equations local and to remove the chemical potential terms. In this case the 
$\lambda_Q$ terms also drop from the simplified equations,%
\footnote{Except possibly for the usual subtleties in the equation for $Y_1$.} 
due to the relation 
$\lambda_{Q-1}(u)\lambda_{Q+1}(u)=\lambda_Q(u-i/2)\lambda_Q(u+i/2)$. The 
derivation of the $Y$-system equations \cite{Gromov:2009tv} then follows as 
usual \cite{AF-TBA}, and they are identical to the regular equations for the spectral 
problem in $AdS$/CFT.

Assuming $Y_Q=0$ one finds the asymptotic solution for the auxiliary particles which is 
identical to the case of twisted periodic TBA \cite{deLeeuw:2012hp}
\beq
\begin{gathered}
Y^{(\pm)\circ}_{w|M}=\frac{\sin M\theta\sin(M+2)\theta}{\sin^2\theta}\,,
\qquad
Y^{(\pm)\circ}_{vw|M}=\frac{\sin M\phi\sin(M+2)\phi}{\sin^2\phi}\,,
\\
Y^{(\pm)\circ}_+=\frac{\cos\phi}{\cos\theta}\,,
\qquad
Y^{(\pm)\circ}_-=\frac{\cos\phi}{\cos\theta}\,,
\end{gathered}
\eeq
Feeding this back into the equation for $Y_Q$ and including the boundary driving term 
gives
\beq
Y^{\circ}_Q=4(\cos\phi-\cos\theta)^2\frac{\sin^2Q\phi}{\sin^2\phi}\,\lambda_Q\,e^{-2(L+1)\tilde E_Q}
\eeq
Note that the assumption of small $Y_Q$ is true only for $u\neq0$ and this is not 
a good approximate solution at $u=0$. Still as shown in the previous section, 
the pole contribution does give the correct one loop gauge theory result.

\section{Discussion}
\label{sec:discuss}

This paper presents the first step in finding the exact quark--antiquark potential in 
$\cN=4$ SYM: The set of twisted boundary TBA equations whose solution is conjectured to give 
the desired function. As explained in the introduction, with the extra two parameters 
$\theta$ and $\phi$ the generalized potential is the same as the generalized 
cusp anomalous dimension.

The resulting quantity gives the conformal dimension of cusps in arbitrary Maldacena-Wilson 
loops in this theory. The focus here has been on the most symmetric cases, 
of the antiparallel lines or infinite cusp. But the divergences are a UV property 
and depend only on the local structure of the Wilson loop. Knowledge of the 
exact generalized cusp anomalous dimension allows therefore to renormalize 
any Wilson loop (with the usual scalar coupling) in this theory 
with arbitrary number of cusps. One follows usual normalization, now that the 
UV behavior near the cusp is under control, see \cite{Brandt:1981kf}.

The discussion here was for real angle $\phi$ as appropriate for 
the Euclidean theory (or a Euclidean cusp in the Lorentzian theory). There is no 
reason to restrict to that. All the expression are analytic in $\phi$ and a Wick-rotation 
$\phi=i\varphi$ gives the cusp anomalous dimension for an arbitrary 
boost angle $\varphi$. At large $\varphi$ the result should be proportional to 
$\varphi$, with the coefficient being a quarter of the universal 
cusp anomalous dimension $\gamma_\text{cusp}/4$. 
In particular for large $\varphi$ the BES 
equation for the cusp anomalous dimension \cite{bes} 
(and its solution \cite{Basso:2007wd}) 
should be recovered from the twisted boundary TBA equation.

For large but finite $\varphi$ this gives the regularized cusp anomalous 
dimension which plays a role in scattering amplitudes in $\cN=4$ SYM when 
imposing a Higgs-VEV regulator as in \cite{Alday:2009zm,Henn:2010bk,CHMS2}.

The TBTBA equations contain much more information than that, though. As usual, 
by contour deformation one can calculate the exact spectrum of excited states 
of the relevant string (or cusped Wilson loops with local operator insertions). For 
large $L$ one can approximate the solution by solving the boundary Bethe-Yang 
equations instead.

In the case of large imaginary angle, these are excitation of the string with 
lightlike cusp describing 
scattering amplitudes in $AdS$. This spectrum is a crucial ingredient in the 
OPE approach to lightlike Wilson loops as discussed in 
\cite{Alday:2010ku,Gaiotto:2010fk} 
and should be closely related to the excitation spectrum of the spinning string 
as studied in \cite{Basso:2010in} (see also \cite{Basso:2011rs}).

There are many other questions left for the future. One is numerical solutions of 
the TBTBA equations, giving the interpolating function for the quark--antiquark 
potential and other interesting quantities in this model.

Another is to try to solve these equations at large coupling and reproduce semiclassical 
string theory results, as was done from the $Y$-system in \cite{Gromov:2009tq}. 
Normally these techniques work only for large $L$ and all other charges also large, 
since otherwise there is no semiclassical string description. In this case, though, 
the classical string description exists already for $L=0$ as the usual description 
of Wilson loops in $AdS_5\times\bS^5$. It should therefore be that the 
algebraic curves describing the string duals of these Wilson loops 
\cite{maldacena-wl,rey-yee,DGO,DGRT-big,Dru-For}
can be derived from these equations.

The same tools used here can be used to study other open spin--chain models 
which arise in the $AdS$/CFT correspondence. The simplest are the D5-brane 
defects, with a very similar symmetry to the Wilson loop. There are two natural 
choices for the vacuum there, one with and one without boundary degrees of 
freedom. The latter should be essentially the same as the model studied here, 
while also the former should not be much different. One would need to 
calculate the boundary scalar factor and then always carry through the two 
boundary excitations, so similar to calculating excited states in this model.

Another interesting system to study is ABJM theory 
\cite{abjm}. The $1/2$ BPS Wilson loop is known in that theory \cite{dt}, but the 
quantity calculated here --- the quark--antiquark potential or cusp anomalous 
dimension has not been calculated in the gauge theory. The leading classical 
result from string theory on $AdS_4\times\bC\bP^4$ is the same as for 
$AdS_5\times\bS^5$ (with the appropriate identification between the string tension 
and the gauge coupling).

\subsection*{Acknowledgements}

I am grateful to 
Valentina Forini for initial collaboration on this project, to
Gleb Arutyunov and Marius de Leeuw for discussions and for sharing computer code with me 
and to 
Changrim Ahn, 
Zoltan Bajnok, 
Ben Doyon, 
Davide Fioravanti, 
Sergei Frolov, 
Davide Gaiotto, 
Fedor Levkovich-Maslyuk, 
Arthur Lipstein, 
Tomasz {\L}ukowski,
Juan Maldacena, 
Carlo Meneghelli,
Rafael Nepomechie,
Vidas Regelskis, 
Amit Sever, 
Dima Volin, 
Gerard Watts
and 
Kostya Zarembo
for illuminating discussions
I would also like to acknowledge the hospitality of the CERN theory group, 
GGI Florence, Perimeter Institute,
The Newton Institute, Nordita and LMU Munich.
This work is underwritten by an advanced fellowship of the 
Science \& Technology Facilities Council.

\appendix
\section{Supersymmetry}
\label{app:susy}

The vacuum of $\cN=4$ SYM is invariant under the $PSU(2,2|4)$ superconformal group, 
summarized here following \cite{Beisert:2004ry}.

Denote by $\gL^{\alpha}_{\ \beta}$, $\bar\gL^{{\dot\alpha}}_{\ {\dot\beta}}$ the 
generators of the $SU(2)_L\times SU(2)_R$ Lorentz group, and by  $\gR^A_{\ B}$ the 15
generators of the $R$-symmetry group $SU(4)$. The remaining bosonic
generators are the translations $\gP^{\dot\beta\alpha}$, the
special conformal transformations $\gK_{\alpha{\dot\beta}}$ and the
dilatation $\gD$. Finally the 32 fermionic generators are the
Poincar\'e supersymmetries $\gQ_A^{\alpha}$, $\bar\gQ^{\dot\alpha A}$ and
the superconformal supersymmetries $\gS_\alpha^A$,
$\bar\gS_{\dot\alpha A}$.

The commutators of any generator $\gG$ with
$\gL^{\alpha}_{\ \beta}$, $\bar \gL^{{\dot\alpha}}_{\ {\dot\beta}}$
and $\gR^A_{\ B}$ are canonically dictated by the index structure
\bal
\big[\gL^{\alpha}_{\ \beta},\gG^\gamma\big]
&=\delta_\beta^\gamma \gG^\alpha
-{\textstyle\frac{1}{2}}\delta^\alpha_\beta \gG^\gamma\,,
&\qquad
\big[\gL^{\alpha}_{\ \beta},\gG_\gamma\big]
&=-\delta^\alpha_\gamma \gG_\beta
+{\textstyle\frac{1}{2}}\delta^\alpha_\beta \gG_\gamma\,,
\\
\big[\bar{\gL}^{{\dot\alpha}}_{\ {\dot\beta}},\gG^{{\dot\gamma}}\big]
&=\delta_{{\dot\beta}}^{{\dot\gamma}} \gG^{{\dot\alpha}}
-{\textstyle\frac{1}{2}}\delta^{{\dot\alpha}}_{{\dot\beta}} \gG^{{\dot\gamma}}\,,
&\qquad
\big[\bar{\gL}^{{\dot\alpha}}_{\ {\dot\beta}},\gG_{{\dot\gamma}}\big]
&=-\delta^{{\dot\alpha}}_{{\dot\gamma}} \gG_{{\dot\beta}}
+{\textstyle\frac{1}{2}}\delta^{{\dot\alpha}}_{{\dot\beta}} \gG_{{\dot\gamma}}\,,
\\
\big[\gR^A_{\ B},\gG^C\big]
&=\delta_B^C \gG^A-{\textstyle\frac{1}{4}}\delta^A_B \gG^C\,,
&\qquad
\big[\gR^A_{\ B},\gG_C\big]
&=-\delta^A_C \gG_B+{\textstyle\frac{1}{4}}\delta^A_B \gG_C\,.
\label{LR}
\eal
while commutators with the
dilatation operator $\gD$ are given by $\big[\gD,\gG \big] =
\text{dim}(\gG)\,\gG$, where $\text{dim}(\gG)$ is the
dimension of the generator $\gG$.

The remaining non-trivial commutators are
\begin{equation}
\begin{aligned}
&\big\{ \gQ_A^\alpha, \bar\gQ^{\dot\beta B} \big\}=
\delta_A^B \gP^{\dot\beta \alpha}\,,
&\qquad
&\big\{\gS^A_\alpha,\bar\gS_{\dot\beta B} \big\}
= \delta_B^A \gK_{\alpha\dot\beta}\,, 
\\
&\big[\gK_{\alpha\dot\beta}, \gQ_A^\gamma \big]=
\delta_\alpha^\gamma\bar\gS_{\dot\beta A}\,,
&\qquad
&\big[\gK_{\alpha\dot\beta}, \bar\gQ^{\dot\gamma A} \big] 
=\delta_{\dot\beta}^{\dot\gamma} \gS_{\alpha}^A\,, 
\\
&\big[\gP^{\dot\alpha\beta}, \gS_\gamma^A \big]
= -\delta^\beta_\gamma\bar\gQ^{\dot\alpha A}\,,
&\qquad
&\big[\gP^{\dot\alpha\beta}, \bar\gS_{\dot\gamma A}\big] 
=-\delta^{\dot\alpha}_{\dot\gamma} \gQ^{\beta}_A\,, 
\\
&\big\{ \gQ_A^\alpha , \gS_\beta^B \big\} 
= \delta_A^B\gL^\alpha_{\ \beta} +\delta_\beta^\alpha \gR^B_{\ A} 
+ {\textstyle\frac{1}{2}} \delta_A^B \delta_\beta^\alpha \gD\,, 
\hskip-2in
\\
&\big\{\bar\gQ^{\dot\alpha A} , \bar\gS_{\dot\beta B} \big\}
= \delta_B^A \bar\gL^{\dot\alpha}_{\ \dot\beta} -\delta_{\dot\beta}^{\dot\alpha} \gR^A_{\ B} 
+ {\textstyle\frac{1}{2}} \delta^A_B \delta_{\dot\beta}^{\dot\alpha} \gD\,, 
\hskip-2in
\\
&\big[ \gK_{\alpha\dot\beta},\gP^{\dot\gamma\delta}\big]
=\delta_{\dot\beta}^{\dot\gamma}\gL^{\delta}_{\ \alpha} 
+ \delta_{\alpha}^{\delta}\bar\gL^{\dot \gamma}_{\ \dot\beta} 
+ \delta_\alpha^\delta\delta^{\dot\gamma}_{\dot\beta} \gD\,.
\hskip-2in
\end{aligned}
\label{bigalgebra1}
\end{equation}

To write down the integrable spin--chain for $\cN=4$ SYM one chooses a vacuum 
corresponding to one complex scalar field, usually labeled $Z=\Phi^5+i\Phi^6$. This 
choice of spin--chain vacuum breaks the symmetry group 
$PSU(2,2|4)\to PSU(2|2)^2$ with supercharges $(Q_\alpha^a,S^\alpha_a)$ and 
$(\bar Q^{\dot\alpha}_{\dot b},\bar S_{\dot\alpha}^{\dot b})$ where 
$a,b\in\{1,2\}$ and $\dot a,\dot b\in\{3,4\}$. Explicitly, the two copies of $\psu(2|2)$ 
are the following subset of the generators of $\psu(2,2|4)$
\bal
\label{su222}
Q_a^\alpha&=\gQ_a^\alpha\,,\qquad
&S^a_\alpha&=\gS^a_\alpha\,,\qquad
&L^\alpha_{\ \beta}&=\gL^\alpha_{\ \beta}\,,\qquad
&R^a_{\ b}&=\gR^a_{\ b}-{\textstyle\frac{1}{2}}\delta^a_b\gR^c_{\ c}\,;
\\
\bar Q_{\dot a}^{\dot\alpha}&=\epsilon_{\dot a\dot b}\bar \gQ^{\dot\alpha\dot b}\,,\qquad
&\bar S^{\dot a}_{\dot\alpha}&=-\epsilon^{\dot a\dot b}\bar \gS_{\dot\alpha\dot b}\,,\qquad
&\bar L^{\dot\alpha}_{\ \dot\beta}&=\bar \gL^{\dot\alpha}_{\ \dot\beta}\,,\qquad
&\bar R^{\dot a}_{\ \dot b}&=\gR^{\dot a}_{\ \dot b}-{\textstyle\frac{1}{2}}\delta^{\dot a}_{\dot b}\gR^{\dot c}_{\ \dot c}\,.
\eal
In the presence of the Wilson loop or domain wall this 
is further broken down to a single copy of $PSU(2|2)$.

\section{The $\psu(2|2)$ spin--chain}
\label{app:magnons}

The spin--chain description of single trace local operators, which we will use also for 
the purposes of studying the boundary changing operators involves choosing a ground 
state, usually taken to be $\Tr Z^J$ with $Z=\Phi^5+i\Phi^6$ and considering the 
excitations about it. The Choice of scalar $Z$ breaks the symmetry group 
$PSU(2,2|4)\to PSU(2|2)^2\times U(1)$. Magnons are therefore classified by 
representations of the broken group. The basic magnons are in the 
fundamental representation of each of the $PSU(2|2)$, so we can first treat the 
magnons as if charged only under one of the groups, remembering to pair them up later.

When constructing magnon excitations on the spin--chain it is useful to consider the 
central extension of the $\psu(2|2)$ algebras. The commutators of 
$L^\alpha_{\ \beta}$, $\bar L^{\dot\alpha}_{\ \dot\beta}$, $R^a_{\ b}$ and 
$R^{\dot a}_{\ \dot b}$ are inherited from \eqn{LR}. The central extension for the 
$Q$ and $S$ commutation relations are
\bal
\{Q^\alpha_a,Q^\beta_b\}&=\epsilon^{\alpha\beta}\epsilon_{ab}P\,,&
\{S_\alpha^a,S_\beta^b\}&=\epsilon_{\alpha\beta}\epsilon^{ab}K\,,
\\
\{Q^\alpha_a,S_\beta^b\}&=\delta_a^b L^\alpha_\beta+\delta^\alpha_\beta R_a^b
+\delta^\alpha_\beta\delta_a^b C\,,
\eal
and likewise for the second copy of $\psu(2|2)$.

The fundamental representation has a pair of bosons $\phi^a$ and a pair of fermions $\psi^\alpha$. 
The algebra acts on them by
\bal
Q_a^\alpha\phi^b&=a\delta_a^b\psi^\alpha\,,&
Q_a^\alpha\psi^\beta&=-\frac{iaf}{x^-}\epsilon^{\alpha\beta}\epsilon_{ab}\phi^b\,,
\\
S^a_\alpha\phi^b&=\frac{ia}{fx^+}\epsilon_{\alpha\beta}\epsilon^{ab}\psi^\beta\,,
&\qquad
S^a_\alpha\psi^\beta&=a\delta_\alpha^\beta\phi^b\,,
\eal
where $a$, the spectral parameters $x^\pm$, the coupling $g$ and the magnon momentum $p$ 
are all related by
\beq
x^++\frac{1}{x^+}-x^--\frac{1}{x^-}=\frac{i}{g}\,,
\qquad
a^2=ig(x^--x^+)\,,
\qquad
e^{ip}=\frac{x^+}{x^-}\,.
\eeq
The central charges for this representation are given by
\beq
C=a^2-\frac{1}{2}\,,
\qquad
P=gf(1-e^{ip})\,,
\qquad
K=gf^{-1}(1-e^{-ip})\,.
\eeq

For a single magnon we can eliminate the parameter $f$ by rescaling 
$Q\to f^{1/2}Q$, $S\to f^{-1/2}S$ and $\psi\to f^{1/2}\psi$. It is crucial though for integrability 
when constructing muli-magnon states. In that case we take for the $k^\text{th}$ magnon 
$f_k=e^{i\sum_{j=k+1}^M p_j}$. It is simple to show by induction that then for $M$ magnons 
the total central charges are
\beq
P=\sum_{k=1}^M P_k=gf_M\big(1-e^{i\sum_{k=1}^Mp_k}\big)\,,
\qquad
K=\sum_{k=1}^M K_k=gf_M^{-1}\big(1-e^{-i\sum_{k=1}^Mp_k}\big)\,.
\eeq

The S-matrix exchanges two fundamental representations and has the general form
\bal
\label{S-matrix1}
s^{12}\phi_1^a\phi_2^b
&=A^{12}\phi_2^{\{a}\phi_1^{b\}}+B^{12}\phi_2^{[a}\phi_1^{b]}
+\textstyle{\frac{1}{2}}C^{12}\epsilon^{ab}\epsilon_{\alpha\beta}\psi_2^\alpha\psi_1^\beta\,,
\\
s^{12}\psi_1^\alpha\psi_2^\beta
&=D^{12}\psi_2^{\{\alpha}\psi_1^{\beta\}}+E^{12}\psi_2^{[\alpha}\psi_1^{\beta]}
+\textstyle{\frac{1}{2}}F^{12}\epsilon^{\alpha\beta}\epsilon_{ab}\phi_2^a\phi_1^b\,,
\\
s^{12}\phi_1^a\psi_2^\beta
&=G^{12}\psi_2^\beta\phi_1^a+H^{12}\phi_2^a\psi_1^\beta\,,
\\
s^{12}\psi_1^\alpha\phi_2^b
&=K^{12}\psi_2^\alpha\phi_1^b+L^{12}\phi_2^b\psi_1^\alpha\,.
\eal
Up to an overall scalar phase, $S_0$, all the terms in this matrix are fixed by 
symmetry \cite{Beisert:2005tm,Beisert:2006qh}. This is achieved by imposing that the 
left and right hand side transform in the same way under $\psu(2|2)$. The solution is
\bal
\label{S-matrix2}
A^{12}&=S_0\frac{x_2^+-x_1^-}{x_2^--x_1^+}\,,
&\qquad
B^{12}&=A^{12}\left(1-2\frac{1-1/x_2^-x_1^+}{1-1/x_2^+x_1^+}\,\frac{x_2^--x_1^-}{x_2^+-x_1^-}\right),
\\
D^{12}&=-S_0\,,
&\qquad
E^{12}&=D^{12}\left(1-2\frac{1-1/x_2^+x_1^-}{1-1/x_2^-x_1^-}\,\frac{x_2^+-x_1^+}{x_2^--x_1^+}\right),
\\
G^{12}&=S_0\frac{x_2^+-x_1^+}{x_2^--x_1^+}\,,
&\qquad
C^{12}&=S_0\frac{2a_1a_2}{x_1^-x_2^-}\frac{f_2}{g}
\frac{1}{1-1/x_2^+x_1^+}\,\frac{x_2^--x_1^-}{x_2^--x_1^+}\,,
\\
L^{12}&=S_0\frac{x_2^--x_1^-}{x_2^--x_1^+}\,,
&\qquad
F^{12}&=S_0\frac{2a_1a_2}{x_1^+x_2^+}\frac{1}{gf_2}
\frac{1}{1-1/x_2^-x_1^-}\,\frac{x_2^+-x_1^+}{x_2^--x_1^+}\,,
\\
H^{12}&=
K^{12}=S_0\frac{ia_1a_2}{g}\frac{1}{x_2^--x_1^+}\,.\hskip-1in
\eal

\section{Reflection matrix}
\label{app:reflect}

The reflection matrix is $\bR(p)=R_0(p)\hat\bS(p,-p)$, where $\bS$ is the scattering matrix 
written above with the factor of $S_0$ removed. $R_0(p)=\sigma_B(p)/\sigma(p,-p)$ 
is the boundary scalar factor discussed in Section~\ref{sec:dressing} and evaluated in 
the next appendix. Under $p\to-p$ the spectral parameters 
transform as $x^\pm\to-x^\mp$. Therefore using the same ansatz as in \eqn{S-matrix1}, 
the coefficients of the reflection matrix are
\bal
\label{R-matrix2}
A_{12}&=R_0\frac{x^-}{x^+}\,,
&\qquad
B^{12}&=-R_0\frac{x^- ((x^+)^3+x^-)}{(x^+)^2 (1+x^+x^-)}\,,
\\
D^{12}&=-R_0\,,
&\qquad
E^{12}&=R_0 \frac{(x^++(x^-)^3)}{x^- (1+x^+x^- )}\,,
\\
G^{12}&=L^{12}=R_0\frac{x^++x^-}{2x^+}\,,
&\qquad
C^{12}&=i R_0\frac{(x^++x^-)(x^+-x^-)}{x^- (1+x^+x^- )}\,,
\\
H^{12}&=
K^{12}=-R_0\frac{x^+-x^-}{2x^+}\,,
&\qquad
F^{12}&=i R_0 \frac{x^-(x^++x^-)(x^+-x^-)}{(x^+)^2(1+x^+x^- )}\,.
\eal

\section{Boundary dressing factor}
\label{app:dress}

\subsection{derivation}
\label{app:dress-derive}

The boundary dressing phase $\sigma_B(u)$ should satisfy the crossing and unitarity 
equations \eqn{crossing}
\beq
\label{crossing-app}
\sigma_B(u)\sigma_B(\bar u)= \frac{x^-+1/x^-}{x^++1/x^+}\,,
\qquad
\sigma_B(u)\sigma_B(- u)=1\,.
\eeq
This equation can be solved the same way the bulk one is \cite{Volin:2009uv} 
(which is significantly simpler than the way it was originally found \cite{Beisert:2006ib,bes}), 
see also \cite{Vieira:2010kb}. 
Take the ansatz
\beq
\sigma_B(u)=e^{i(\chi_B(x^+)-\chi_B(x^-))}\,,
\eeq
and defining $\tilde\sigma_B(x)=e^{i(\chi_B(x)+\chi_B(1/x))}$, then using that 
under crossing transformations $x^\pm\to1/x^\pm$ the crossing equation 
\eqn{crossing-app} can be written as
\beq
\sigma_B(u)\sigma_B(\bar u)=\tilde\sigma_B(x^+)\tilde\sigma_B(x^-)= \frac{x^-+1/x^-}{x^++1/x^+}
=\frac{u-i/2}{u+i/2}\,.
\eeq
In terms of the shift operators 
$D^\pm=e^{\pm\frac{i}{2}\partial_u}$ this equation becomes
\beq
\label{tilsig}
\tilde\sigma_B(x)^{D^+-D^-}= (x+1/x)^{D^--D^+}=u^{D^--D^+}\,.
\eeq
So
\beq
\tilde\sigma_B(x)=(x+1/x)^{F[D]}\,,
\qquad F[D]\sim\frac{D^--D^+}{D^+-D^-}
\eeq
Taking $F[D]=-1$ will not lead to the desired answer, rather will introduce the false 
analytic structure. Instead one can try
\beq
F[D]=\frac{D^{-2}}{1-D^{-2}}+\frac{D^{+2}}{1-D^{+2}}
=\sum_{n=1}^\infty\left(D^{-2n}+D^{+2n}\right)
\eeq
Therefore%
\footnote{The product is divergent, but this is the natural ragularization of it.}
\beq
\tilde\sigma_B(u)=\prod_{n=1}^\infty (u+in)(u-in)
=\frac{\sinh \pi u}{\pi u}
\eeq
This expression does not quite work. The shifts by $\pm\frac{i}{2}$ gives
\beq
\tilde\sigma_B(u)^{D^+-D^-}=\frac{\tilde\sigma(u+i/2)}{\tilde\sigma(u-i/2)}
=\frac{\sinh \pi (u+i/2)}{\sinh\pi(u-i/2)}\frac{u-i/2}{u+i/2}
=-(x+1/x)^{D^--D^+}
\eeq
This differs from \eqn{tilsig} by a sign, which can be fixed by taking
\beq
\label{discontinuity0}
\tilde\sigma_B(u)=\frac{\sinh 2\pi u}{2\pi u}
\eeq
This corresponds to the choice
\beq
F[D]=\frac{D^{-}}{1-D^{-}}+\frac{D^{+}}{1-D^{+}}
=\sum_{n=1}^\infty\left(D^{-n}+D^{+n}\right)
\eeq
Of course one could replace the $2$ in \eqn{discontinuity0} with any even integer, but this 
would introduce extra poles.

Under crossing transformation $x\to1/x$ so we can interpret the equation for 
$\tilde\sigma_B$ as the 
discontinuity of $\chi_B$ across the cut in the $u$ plane between $\pm2g$. Therefore
\beq
\chi_B(u)=\int_{-2g+i0}^{2g+0i} \frac{dw}{2\pi i}\frac{x(u)-1/x(u)}{x(w)-1/x(w)}\,
\frac{1}{u-w}\,\frac{1}{i}\log \frac{\sinh 2\pi w}{2\pi w}\,.
\eeq
Switching to the $x$ coordinate and ignoring an irrelevant constant gives
\beq
\label{dress}
\chi_B(x)=-i\oint\frac{dy}{2\pi i}\,
\frac{1}{x-y}\log \frac{\sinh 2\pi g(y+1/y)}{2\pi g(y+1/y)}\,.
\eeq

The function $e^{-i\chi(x(u))}$ has a cut for $u\in[-2g,2g]$ in the sheet where 
$|x(u)|>1$. Crossing the cut gives $|x(u)|<1$ where
\beq
\label{discontinuity}
e^{-i\chi(x(u))}
=e^{i\chi(1/x(u))-i\Phi_0}\frac{2\pi u}{\sinh 2\pi u}
\eeq
where $\Phi_0$ is equal to the integral in \eqn{dress} evaluated at $x=0$.
On this sheet, in addition to the cut, this function has poles for all half integer imaginary 
$u$, except for $u=0$.

\subsection{Expansions}
\label{app:dress-expand}

As is done with the bulk dressing factor, we can expand the boundary one for large $x$ as%
\footnote{One can readily check that only even powers will appear in the expansion.}
\beq
i\chi_B(x)=\sum_{r=1}^\infty\frac{c_r(g)}{x^{2r}}
\eeq
Expanding \eqn{dress} at large $x$ are using $z=e^{i\psi}$ we find the integral expression
\beq
c_r(g)=\int_0^{2\pi}\frac{d\psi}{2\pi}\,e^{2ir\psi}\log\frac{\sinh4\pi g\cos\psi}{4\pi g\cos\psi}
\eeq
This integral can also be written as
\beq
c_r(g)=2(-1)^{r+1}\int_0^\infty\frac{J_{2r}(4 g t)}{t(e^t-1)}\,dt
\eeq

One can then expand at weak coupling
\beq
c_r(g)=\sum_{n=0}^\infty c_r^{(n)} g^{2n+2r}
\eeq
and the explicit factors are
\beq
c_r^{(n)}=-\frac{(-4)^{n+r}}{n+r}\frac{(2n+2r)!}{n!(n+2r)!}\zeta(2n+2r)
\eeq

There is also an asymptotic strong coupling expansion
\beq
c_r(g)=\sum_{n=0}^\infty d_r^{(n)} g^{1-n}
\eeq
Apart for the linear and constant terms there are only odd inverse powers of $g$ 
and the explicit factors are
\bal
d_r^{(0)}&=\frac{8(-1)^{r+1}}{4r^2-1}
\\
d_r^{(1)}&=\frac{(-1)^{r}}{2r}
\\
d_r^{(2n)}&=-\frac{16}{(4\pi)^{2n+1}}\,\textstyle\Gamma(n+r-\frac{1}{2})\Gamma(n-r-\frac{1}{2})\zeta(2n)
\qquad n=1,2,\cdots
\\
d_r^{(2n+1)}&=0
\qquad n=1,2,\cdots
\eal
One can now resum the series and express $\chi_B$ as
\beq
i\chi_B(x)=\sum_{n=0}^\infty d^{(n)}(x)\,g^{1-n}
\eeq
where
\begin{align}
d^{(0)}(x)&=-4+4\left(x+\frac{1}{x}\right) \arccot x
\nonumber\\
\label{ds}
d^{(1)}(x)&=-\frac{1}{2} \log\left(1+\frac{1}{x^2}\right)
\\
\nonumber
d^{(2n)}(x)&=-\frac{16}{(4\pi)^{2n+1}}
{\textstyle \Gamma(n+\frac{1}{2})\Gamma(n-\frac{3}{2}) \zeta(2 n)\,
{}_2F_1(n+\frac{1}{2},1,\frac{5}{2}-n,-\frac{1}{x^2})}
\qquad
n=1,2,\cdots
\end{align}
The hypergeometric function is a rational function of $x^2$.

\bibliography{refs}

\providecommand{\href}[2]{#2}\begingroup\raggedright\begin{thebibliography}{10}
\addtolength{\parskip}{-1ex}

\bibitem{wilson}
K.~G. Wilson, ``{Confinement of quarks},''
\href{http://dx.doi.org/10.1103/PhysRevD.10.2445}{{\em Phys. Rev.} {\bf D10}
  (1974)  2445--2459}.

\bibitem{Korchemsky:1988si}
G.~P. Korchemsky, ``{Asymptotics of the Altarelli-Parisi-Lipatov evolution
  kernels of parton distributions},''
\href{http://dx.doi.org/10.1142/S0217732389001453}{{\em Mod. Phys. Lett.} {\bf
  A4} (1989)  1257--1276}.

\bibitem{Korchemsky:1992xv}
G.~Korchemsky and G.~Marchesini, ``{Structure function for large $x$ and
  renormalization of Wilson loop},''
  \href{http://dx.doi.org/10.1016/0550-3213(93)90167-N}{{\em Nucl.Phys.} {\bf
  B406} (1993)  225--258},
\href{http://arxiv.org/abs/hep-ph/9210281}{{\tt hep-ph/9210281}}.

\bibitem{Bern:2005iz}
Z.~Bern, L.~J. Dixon, and V.~A. Smirnov, ``{Iteration of planar amplitudes in
  maximally supersymmetric Yang-Mills theory at three loops and beyond},''
  \href{http://dx.doi.org/10.1103/PhysRevD.72.085001}{{\em Phys.Rev.} {\bf D72}
  (2005)  085001},
\href{http://arxiv.org/abs/hep-th/0505205}{{\tt hep-th/0505205}}.

\bibitem{am1}
L.~F. Alday and J.~M. Maldacena, ``{Gluon scattering amplitudes at strong
  coupling},'' {\em JHEP} {\bf 06} (2007)  064,
\href{http://arxiv.org/abs/0705.0303}{{\tt arXiv:0705.0303}}.

\bibitem{Dru-For}
N.~Drukker and V.~Forini, ``{Generalized quark-antiquark potential at weak and
  strong coupling},'' \href{http://arxiv.org/abs/1105.5144}{{\tt
  arXiv:1105.5144}}.

\bibitem{CHMS2}
D.~Correa, J.~Henn, J.~Maldacena, and A.~Sever, ``{The cusp anomalous dimension
  at three loops and beyond},''
\href{http://arxiv.org/abs/1203.1019}{{\tt arXiv:1203.1019}}.

\bibitem{CHMS1}
D.~Correa, J.~Henn, J.~Maldacena, and A.~Sever, ``{An exact formula for the
  radiation of a moving quark in ${\cal N}=4$ super Yang Mills},''
\href{http://arxiv.org/abs/1202.4455}{{\tt arXiv:1202.4455}}.

\bibitem{zarembo}
K.~Zarembo, ``{Supersymmetric Wilson loops},''
  \href{http://dx.doi.org/10.1016/S0550-3213(02)00693-4}{{\em Nucl. Phys.} {\bf
  B643} (2002)  157--171},
\href{http://arxiv.org/abs/hep-th/0205160}{{\tt hep-th/0205160}}.

\bibitem{Polyakov:1980ca}
A.~M. Polyakov, ``{Gauge fields as rings of glue},''
\href{http://dx.doi.org/10.1016/0550-3213(80)90507-6}{{\em Nucl.Phys.} {\bf
  B164} (1980)  171--188}.

\bibitem{Brandt:1981kf}
R.~A. Brandt, F.~Neri, and M.~Sato, ``{Renormalization of loop functions for
  all loops},'' \href{http://dx.doi.org/10.1103/PhysRevD.24.879}{{\em
  Phys.Rev.} {\bf D24} (1981)  879}.

\bibitem{Makeenko:1979pb}
Y.~Makeenko and A.~A. Migdal, ``{Exact equation for the loop average in
  multicolor QCD},''
\href{http://dx.doi.org/10.1016/0370-2693(79)90131-X,
  10.1016/0370-2693(79)90131-X}{{\em Phys.Lett.} {\bf B88} (1979)  135}.

\bibitem{maldacena-wl}
J.~M. Maldacena, ``{Wilson loops in large $N$ field theories},''
  \href{http://dx.doi.org/10.1103/PhysRevLett.80.4859}{{\em Phys. Rev. Lett.}
  {\bf 80} (1998)  4859--4862},
\href{http://arxiv.org/abs/hep-th/9803002}{{\tt hep-th/9803002}}.

\bibitem{rey-yee}
S.-J. Rey and J.-T. Yee, ``{Macroscopic strings as heavy quarks in large $N$
  gauge theory and anti-de Sitter supergravity},''
  \href{http://dx.doi.org/10.1007/s100520100799}{{\em Eur. Phys. J.} {\bf C22}
  (2001)  379--394},
\href{http://arxiv.org/abs/hep-th/9803001}{{\tt hep-th/9803001}}.

\bibitem{DGO}
N.~Drukker, D.~J. Gross, and H.~Ooguri, ``{Wilson loops and minimal
  surfaces},'' \href{http://dx.doi.org/10.1103/PhysRevD.60.125006}{{\em Phys.
  Rev.} {\bf D60} (1999)  125006},
\href{http://arxiv.org/abs/hep-th/9904191}{{\tt hep-th/9904191}}.

\bibitem{DGRT-big}
N.~Drukker, S.~Giombi, R.~Ricci, and D.~Trancanelli, ``{Supersymmetric Wilson
  loops on $S^3$},''
  \href{http://dx.doi.org/10.1088/1126-6708/2008/05/017}{{\em JHEP} {\bf 05}
  (2008)  017},
\href{http://arxiv.org/abs/0711.3226}{{\tt arXiv:0711.3226}}.

\bibitem{Bena:2003wd}
I.~Bena, J.~Polchinski, and R.~Roiban, ``{Hidden symmetries of the $AdS_5\times
  S^5$ superstring},'' \href{http://dx.doi.org/10.1103/PhysRevD.69.046002}{{\em
  Phys.Rev.} {\bf D69} (2004)  046002},
\href{http://arxiv.org/abs/hep-th/0305116}{{\tt hep-th/0305116}}.

\bibitem{DGT}
N.~Drukker, D.~J. Gross, and A.~A. Tseytlin, ``{Green-Schwarz string in
  $AdS_5\times S^5$: Semiclassical partition function},''
  \href{http://dx.doi.org/10.1088/1126-6708/2000/04/021}{{\em JHEP} {\bf 04}
  (2000)  021},
\href{http://arxiv.org/abs/hep-th/0001204}{{\tt hep-th/0001204}}.

\bibitem{chr}
S.-x. Chu, D.~Hou, and H.-c. Ren, ``{The subleading term of the strong coupling
  expansion of the heavy-quark potential in a ${\mathcal N}=4$ super Yang-Mills
  vacuum},'' \href{http://dx.doi.org/10.1088/1126-6708/2009/08/004}{{\em JHEP}
  {\bf 08} (2009)  004},
\href{http://arxiv.org/abs/0905.1874}{{\tt arXiv:0905.1874}}.

\bibitem{vali-lines}
V.~Forini, ``{Quark-antiquark potential in $AdS$ at one loop},''
  \href{http://dx.doi.org/10.1007/JHEP11(2010)079}{{\em JHEP} {\bf 1011} (2010)
   079}, \href{http://arxiv.org/abs/arXiv:1009.3939}{{\tt
  arXiv:arXiv:1009.3939}}.

\bibitem{dk-spinchain}
N.~Drukker and S.~Kawamoto, ``{Small deformations of supersymmetric Wilson
  loops and open spin-chains},''
  \href{http://dx.doi.org/10.1088/1126-6708/2006/07/024}{{\em JHEP} {\bf 07}
  (2006)  024},
\href{http://arxiv.org/abs/hep-th/0604124}{{\tt hep-th/0604124}}.

\bibitem{df-int}
N.~Drukker and B.~Fiol, ``{On the integrability of Wilson loops in $AdS_5\times
  S^5$: Some periodic ans\"atze},''
  \href{http://dx.doi.org/10.1088/1126-6708/2006/01/056}{{\em JHEP} {\bf 01}
  (2006)  056},
\href{http://arxiv.org/abs/hep-th/0506058}{{\tt hep-th/0506058}}.

\bibitem{CMS}
D.~Correa, J.~Maldacena, and A.~Sever, ``{The quark anti-quark potential and
  the cusp anomalous dimension from a TBA equation},''
\href{http://arxiv.org/abs/1203.1913}{{\tt arXiv:1203.1913}}.

\bibitem{Minahan:2002ve}
J.~Minahan and K.~Zarembo, ``{The Bethe ansatz for ${\cal N}=4$
  superYang-Mills},'' {\em JHEP} {\bf 0303} (2003)  013,
  \href{http://arxiv.org/abs/hep-th/0212208}{{\tt hep-th/0212208}}.

\bibitem{bmn}
D.~E. Berenstein, J.~M. Maldacena, and H.~S. Nastase, ``{Strings in flat space
  and pp waves from ${\cal N}=4$ Super Yang Mills},'' {\em JHEP} {\bf 0204}
  (2002)  013,
\href{http://arxiv.org/abs/hep-th/0202021}{{\tt hep-th/0202021}}.

\bibitem{Berenstein:2005vf}
D.~Berenstein and S.~E. Vazquez, ``{Integrable open spin chains from giant
  gravitons},'' \href{http://dx.doi.org/10.1088/1126-6708/2005/06/059}{{\em
  JHEP} {\bf 0506} (2005)  059},
\href{http://arxiv.org/abs/hep-th/0501078}{{\tt hep-th/0501078}}.

\bibitem{Mann:2006rh}
N.~Mann and S.~E. Vazquez, ``{Classical open string integrability},''
  \href{http://dx.doi.org/10.1088/1126-6708/2007/04/065}{{\em JHEP} {\bf 0704}
  (2007)  065},
\href{http://arxiv.org/abs/hep-th/0612038}{{\tt hep-th/0612038}}.

\bibitem{Hofman:2007xp}
D.~M. Hofman and J.~M. Maldacena, ``{Reflecting magnons},''
  \href{http://dx.doi.org/10.1088/1126-6708/2007/11/063}{{\em JHEP} {\bf 0711}
  (2007)  063}, \href{http://arxiv.org/abs/0708.2272}{{\tt arXiv:0708.2272}}.

\bibitem{DeWolfe:2004zt}
O.~DeWolfe and N.~Mann, ``{Integrable open spin chains in defect conformal
  field theory},'' \href{http://dx.doi.org/10.1088/1126-6708/2004/04/035}{{\em
  JHEP} {\bf 0404} (2004)  035},
\href{http://arxiv.org/abs/hep-th/0401041}{{\tt hep-th/0401041}}.

\bibitem{Correa:2008av}
D.~Correa and C.~Young, ``{Reflecting magnons from D7 and D5 branes},''
  \href{http://dx.doi.org/10.1088/1751-8113/41/45/455401}{{\em J.Phys.A} {\bf
  A41} (2008)  455401}, \href{http://arxiv.org/abs/0808.0452}{{\tt
  arXiv:0808.0452}}.

\bibitem{CRY}
D.~H. Correa, V.~Regelskis, and C.~A. Young, ``{Integrable achiral D5-brane
  reflections and asymptotic Bethe equations},''
  \href{http://dx.doi.org/10.1088/1751-8113/44/32/325403}{{\em J.Phys.A} {\bf
  A44} (2011)  325403},
\href{http://arxiv.org/abs/1105.3707}{{\tt arXiv:1105.3707}}.

\bibitem{GMU}
D.~Gaiotto and J.~Maldacena. Unpublished.

\bibitem{Dekel:2011ja}
A.~Dekel and Y.~Oz, ``{Integrability of Green-Schwarz sigma models with
  boundaries},'' \href{http://dx.doi.org/10.1007/JHEP08(2011)004}{{\em JHEP}
  {\bf 1108} (2011)  004},
\href{http://arxiv.org/abs/1106.3446}{{\tt arXiv:1106.3446}}.

\bibitem{Luscher:1985dn}
M.~{L\"uscher}, ``{Volume dependence of the energy spectrum in massive quantum
  field theories. 1. Stable particle states},''
  \href{http://dx.doi.org/10.1007/BF01211589}{{\em Commun.Math.Phys.} {\bf 104}
  (1986)  177}.

\bibitem{bajnok-janik}
Z.~Bajnok and R.~A. Janik, ``{Four-loop perturbative Konishi from strings and
  finite size effects for multiparticle states},''
  \href{http://dx.doi.org/10.1016/j.nuclphysb.2008.08.020}{{\em Nucl.Phys.}
  {\bf B807} (2009)  625--650}, \href{http://arxiv.org/abs/0807.0399}{{\tt
  arXiv:0807.0399}}.

\bibitem{mos}
Y.~Makeenko, P.~Olesen, and G.~W. Semenoff, ``{Cusped SYM Wilson loop at two
  loops and beyond},''
  \href{http://dx.doi.org/10.1016/j.nuclphysb.2006.05.002}{{\em Nucl. Phys.}
  {\bf B748} (2006)  170--199},
\href{http://arxiv.org/abs/hep-th/0602100}{{\tt hep-th/0602100}}.

\bibitem{kr-wl}
G.~P. Korchemsky and A.~V. Radyushkin, ``{Renormalization of the Wilson loops
  beyond the leading order},''
\href{http://dx.doi.org/10.1016/0550-3213(87)90277-X}{{\em Nucl. Phys.} {\bf
  B283} (1987)  342--364}.

\bibitem{Kotikov:2003fb}
A.~V. Kotikov, L.~N. Lipatov, and V.~N. Velizhanin, ``{Anomalous dimensions of
  Wilson operators in ${\cal N} = 4$ SYM theory},''
  \href{http://dx.doi.org/10.1016/S0370-2693(03)00184-9}{{\em Phys. Lett.} {\bf
  B557} (2003)  114--120},
\href{http://arxiv.org/abs/hep-ph/0301021}{{\tt hep-ph/0301021}}.

\bibitem{Beisert:2004ry}
N.~Beisert, ``{The dilatation operator of ${\cal N} = 4$ super Yang-Mills
  theory and integrability},''
  \href{http://dx.doi.org/10.1016/j.physrep.2004.09.007}{{\em Phys. Rept.} {\bf
  405} (2005)  1--202},
\href{http://arxiv.org/abs/hep-th/0407277}{{\tt hep-th/0407277}}.

\bibitem{Janik-cross}
R.~A. Janik, ``{The $AdS_5 \times S^5$ superstring worldsheet S-matrix and
  crossing symmetry},''
  \href{http://dx.doi.org/10.1103/PhysRevD.73.086006}{{\em Phys.Rev.} {\bf D73}
  (2006)  086006},
\href{http://arxiv.org/abs/hep-th/0603038}{{\tt hep-th/0603038}}.

\bibitem{Ahn:2007bq}
C.~Ahn, D.~Bak, and S.-J. Rey, ``{Reflecting magnon bound states},''
  \href{http://dx.doi.org/10.1088/1126-6708/2008/04/050}{{\em JHEP} {\bf 0804}
  (2008)  050},
\href{http://arxiv.org/abs/0712.4144}{{\tt arXiv:0712.4144}}.

\bibitem{Volin:2009uv}
D.~Volin, ``{Minimal solution of the $AdS$/CFT crossing equation},''
  \href{http://dx.doi.org/10.1088/1751-8113/42/37/372001}{{\em J.Phys.A} {\bf
  A42} (2009)  372001},
\href{http://arxiv.org/abs/0904.4929}{{\tt 0904.4929}}.

\bibitem{Arutyunov:2004vx}
G.~Arutyunov, S.~Frolov, and M.~Staudacher, ``{Bethe ansatz for quantum
  strings},'' \href{http://dx.doi.org/10.1088/1126-6708/2004/10/016}{{\em JHEP}
  {\bf 0410} (2004)  016},
\href{http://arxiv.org/abs/hep-th/0406256}{{\tt hep-th/0406256}}.

\bibitem{Aharony:2008an}
O.~Aharony and D.~Kutasov, ``{Holographic duals of long open strings},''
  \href{http://dx.doi.org/10.1103/PhysRevD.78.026005}{{\em Phys.Rev.} {\bf D78}
  (2008)  026005},
\href{http://arxiv.org/abs/0803.3547}{{\tt arXiv:0803.3547}}.

\bibitem{Beisert:2005if}
N.~Beisert and R.~Roiban, ``{Beauty and the twist: The Bethe ansatz for twisted
  ${\cal N}=4$ SYM},''
  \href{http://dx.doi.org/10.1088/1126-6708/2005/08/039}{{\em JHEP} {\bf 0508}
  (2005)  039}, \href{http://arxiv.org/abs/hep-th/0505187}{{\tt
  hep-th/0505187}}.

\bibitem{Ahn:2010yv}
C.~Ahn, Z.~Bajnok, D.~Bombardelli, and R.~I. Nepomechie, ``{Finite-size effect
  for four-loop Konishi of the $\beta$-deformed ${\cal N}=4$ SYM},''
  \href{http://dx.doi.org/10.1016/j.physletb.2010.08.056}{{\em Phys.Lett.} {\bf
  B693} (2010)  380--385}, \href{http://arxiv.org/abs/1006.2209}{{\tt
  arXiv:1006.2209}}.

\bibitem{Gromov:2010dy}
N.~Gromov and F.~Levkovich-Maslyuk, ``{$Y$-system and $\beta$-deformed ${\cal
  N}=4$ Super-Yang-Mills},''
  \href{http://dx.doi.org/10.1088/1751-8113/44/1/015402}{{\em J.Phys.A} {\bf
  A44} (2011)  015402}, \href{http://arxiv.org/abs/1006.5438}{{\tt
  arXiv:1006.5438}}.

\bibitem{Arutyunov:2010gu}
G.~Arutyunov, M.~de~Leeuw, and S.~J. van Tongeren, ``{Twisting the mirror
  TBA},'' \href{http://dx.doi.org/10.1007/JHEP02(2011)025}{{\em JHEP} {\bf
  1102} (2011)  025},
\href{http://arxiv.org/abs/1009.4118}{{\tt arXiv:1009.4118}}.

\bibitem{Ahn:2011xq}
C.~Ahn, Z.~Bajnok, D.~Bombardelli, and R.~I. Nepomechie, ``{TBA, NLO Luscher
  correction, and double wrapping in twisted $AdS$/CFT},''
  \href{http://dx.doi.org/10.1007/JHEP12(2011)059}{{\em JHEP} {\bf 1112} (2011)
   059},
\href{http://arxiv.org/abs/1108.4914}{{\tt arXiv:1108.4914}}.

\bibitem{deLeeuw:2012hp}
M.~de~Leeuw and S.~J. van Tongeren, ``{The spectral problem for strings on
  twisted $AdS_5 \times S^5$},''
\href{http://arxiv.org/abs/1201.1451}{{\tt arXiv:1201.1451}}.

\bibitem{Arutyunov:2007tc}
G.~Arutyunov and S.~Frolov, ``{On string S-matrix, bound states and TBA},''
  \href{http://dx.doi.org/10.1088/1126-6708/2007/12/024}{{\em JHEP} {\bf 0712}
  (2007)  024},
\href{http://arxiv.org/abs/0710.1568}{{\tt arXiv:0710.1568}}.

\bibitem{Cherednik:1985vs}
I.~Cherednik, ``{Factorizing particles on a half line and root systems},''
\href{http://dx.doi.org/10.1007/BF01038545, 10.1007/BF01038545}{{\em
  Theor.Math.Phys.} {\bf 61} (1984)  977--983}.

\bibitem{Sklyanin:1988yz}
E.~Sklyanin, ``{Boundary conditions for integrable quantum systems},''
\href{http://dx.doi.org/10.1088/0305-4470/21/10/015}{{\em J.Phys.A} {\bf A21}
  (1988)  2375}.

\bibitem{Murgan:2008fs}
R.~Murgan and R.~I. Nepomechie, ``{Open-chain transfer matrices for
  $AdS$/CFT},'' \href{http://dx.doi.org/10.1088/1126-6708/2008/09/085}{{\em
  JHEP} {\bf 0809} (2008)  085},
\href{http://arxiv.org/abs/0808.2629}{{\tt arXiv:0808.2629}}.

\bibitem{Galleas:2009ye}
W.~Galleas, ``{The Bethe ansatz equations for reflecting magnons},''
  \href{http://dx.doi.org/10.1016/j.nuclphysb.2009.04.024}{{\em Nucl.Phys.}
  {\bf B820} (2009)  664--681},
\href{http://arxiv.org/abs/0902.1681}{{\tt arXiv:0902.1681}}.

\bibitem{Correa:2009mz}
D.~Correa and C.~Young, ``{Finite size corrections for open strings/open chains
  in planar $AdS$/CFT},''
  \href{http://dx.doi.org/10.1088/1126-6708/2009/08/097}{{\em JHEP} {\bf 0908}
  (2009)  097}, \href{http://arxiv.org/abs/0905.1700}{{\tt arXiv:0905.1700}}.

\bibitem{Bajnok:2010ui}
Z.~Bajnok and L.~Palla, ``{Boundary finite size corrections for multiparticle
  states and planar $AdS$/CFT},''
  \href{http://dx.doi.org/10.1007/JHEP01(2011)011}{{\em JHEP} {\bf 1101} (2011)
   011},
\href{http://arxiv.org/abs/1010.5617}{{\tt arXiv:1010.5617}}.

\bibitem{Ahn:2010ws}
C.~Ahn, Z.~Bajnok, D.~Bombardelli, and R.~I. Nepomechie, ``{Twisted Bethe
  equations from a twisted S-matrix},''
  \href{http://dx.doi.org/10.1007/JHEP02(2011)027}{{\em JHEP} {\bf 1102} (2011)
   027}, \href{http://arxiv.org/abs/1010.3229}{{\tt arXiv:1010.3229 [hep-th]}}.

\bibitem{Bajnok:2004tq}
Z.~Bajnok, L.~Palla, and G.~Takacs, ``{Finite size effects in quantum field
  theories with boundary from scattering data},''
  \href{http://dx.doi.org/10.1016/j.nuclphysb.2005.03.021}{{\em Nucl.Phys.}
  {\bf B716} (2005)  519--542},
\href{http://arxiv.org/abs/hep-th/0412192}{{\tt hep-th/0412192}}.

\bibitem{Bajnok:2006dn}
Z.~Bajnok, L.~Palla, and G.~Takacs, ``{Boundary one-point function, Casimir
  energy and boundary state formalism in $D+1$ dimensional QFT},''
  \href{http://dx.doi.org/10.1016/j.nuclphysb.2007.02.023}{{\em Nucl.Phys.}
  {\bf B772} (2007)  290--322},
\href{http://arxiv.org/abs/hep-th/0611176}{{\tt hep-th/0611176}}.

\bibitem{LeClair:1995uf}
A.~LeClair, G.~Mussardo, H.~Saleur, and S.~Skorik, ``{Boundary energy and
  boundary states in integrable quantum field theories},''
  \href{http://dx.doi.org/10.1016/0550-3213(95)00435-U}{{\em Nucl.Phys.} {\bf
  B453} (1995)  581--618}, \href{http://arxiv.org/abs/hep-th/9503227}{{\tt
  hep-th/9503227}}.

\bibitem{Z-TBA}
A.~Zamolodchikov, ``{Thermodynamic Bethe Ansatz in relativistic models. Scaling
  three state Potts and Lee-Yang models},''
  \href{http://dx.doi.org/10.1016/0550-3213(90)90333-9}{{\em Nucl.Phys.} {\bf
  B342} (1990)  695--720}.

\bibitem{Dorey:2000eh}
P.~Dorey, M.~Pillin, R.~Tateo, and G.~Watts, ``{One point functions in
  perturbed boundary conformal field theories},''
  \href{http://dx.doi.org/10.1016/S0550-3213(00)00622-2}{{\em Nucl.Phys.} {\bf
  B594} (2001)  625--659},
\href{http://arxiv.org/abs/hep-th/0007077}{{\tt hep-th/0007077}}.

\bibitem{Gho-Z}
S.~Ghoshal and A.~B. Zamolodchikov, ``{Boundary S matrix and boundary state in
  two-dimensional integrable quantum field theory},''
  \href{http://dx.doi.org/10.1142/S0217751X94001552,
  10.1142/S0217751X94001552}{{\em Int.J.Mod.Phys.} {\bf A9} (1994)
  3841--3886},
\href{http://arxiv.org/abs/hep-th/9306002}{{\tt hep-th/9306002}}.

\bibitem{Bombardelli:2009ns}
D.~Bombardelli, D.~Fioravanti, and R.~Tateo, ``{Thermodynamic Bethe ansatz for
  planar $AdS$/CFT: A proposal},''
  \href{http://dx.doi.org/10.1088/1751-8113/42/37/375401,
  10.1088/1751-8113/42/37/375401}{{\em J.Phys.A} {\bf A42} (2009)  375401},
\href{http://arxiv.org/abs/0902.3930}{{\tt arXiv:0902.3930}}.

\bibitem{Gromov:2009bc}
N.~Gromov, V.~Kazakov, A.~Kozak, and P.~Vieira, ``{Exact spectrum of anomalous
  dimensions of planar ${\cal N} = 4$ supersymmetric Yang-Mills theory: TBA and
  excited states},'' \href{http://dx.doi.org/10.1007/s11005-010-0374-8}{{\em
  Lett.Math.Phys.} {\bf 91} (2010)  265--287},
\href{http://arxiv.org/abs/0902.4458}{{\tt arXiv:0902.4458}}.

\bibitem{AF-TBA}
G.~Arutyunov and S.~Frolov, ``{Thermodynamic Bethe ansatz for the $AdS_5 \times
  S^5$ mirror model},''
  \href{http://dx.doi.org/10.1088/1126-6708/2009/05/068}{{\em JHEP} {\bf 0905}
  (2009)  068},
\href{http://arxiv.org/abs/0903.0141}{{\tt arXiv:0903.0141}}.

\bibitem{Arutyunov:2009ax}
G.~Arutyunov, S.~Frolov, and R.~Suzuki, ``{Exploring the mirror TBA},''
  \href{http://dx.doi.org/10.1007/JHEP05(2010)031}{{\em JHEP} {\bf 1005} (2010)
   031},
\href{http://arxiv.org/abs/0911.2224}{{\tt arXiv:0911.2224}}.

\bibitem{Gromov:2009tv}
N.~Gromov, V.~Kazakov, and P.~Vieira, ``{Exact spectrum of anomalous dimensions
  of planar ${\cal N}=4$ supersymmetric Yang-Mills theory},''
  \href{http://dx.doi.org/10.1103/PhysRevLett.103.131601}{{\em Phys.Rev.Lett.}
  {\bf 103} (2009)  131601},
\href{http://arxiv.org/abs/0901.3753}{{\tt arXiv:0901.3753}}.

\bibitem{bes}
N.~Beisert, B.~Eden, and M.~Staudacher, ``{Transcendentality and crossing},''
  {\em J. Stat. Mech.} {\bf 0701} (2007)  P021,
\href{http://arxiv.org/abs/hep-th/0610251}{{\tt hep-th/0610251}}.

\bibitem{Basso:2007wd}
B.~Basso, G.~Korchemsky, and J.~Kotanski, ``{Cusp anomalous dimension in
  maximally supersymmetric Yang-Mills theory at strong coupling},''
  \href{http://dx.doi.org/10.1103/PhysRevLett.100.091601}{{\em Phys.Rev.Lett.}
  {\bf 100} (2008)  091601},
\href{http://arxiv.org/abs/0708.3933}{{\tt arXiv:0708.3933}}.

\bibitem{Alday:2009zm}
L.~F. Alday, J.~M. Henn, J.~Plefka, and T.~Schuster, ``{Scattering into the
  fifth dimension of ${\cal N}=4$ super Yang-Mills},''
  \href{http://dx.doi.org/10.1007/JHEP01(2010)077}{{\em JHEP} {\bf 1001} (2010)
   077},
\href{http://arxiv.org/abs/0908.0684}{{\tt arXiv:0908.0684}}.

\bibitem{Henn:2010bk}
J.~M. Henn, S.~G. Naculich, H.~J. Schnitzer, and M.~Spradlin,
  ``{Higgs-regularized three-loop four-gluon amplitude in ${\cal N}=4$ SYM:
  Exponentiation and Regge limits},''
  \href{http://dx.doi.org/10.1007/JHEP04(2010)038}{{\em JHEP} {\bf 1004} (2010)
   038},
\href{http://arxiv.org/abs/1001.1358}{{\tt arXiv:1001.1358}}.

\bibitem{Alday:2010ku}
L.~F. Alday, D.~Gaiotto, J.~Maldacena, A.~Sever, and P.~Vieira, ``{An operator
  product expansion for polygonal null Wilson loops},''
  \href{http://dx.doi.org/10.1007/JHEP04(2011)088}{{\em JHEP} {\bf 1104} (2011)
   088},
\href{http://arxiv.org/abs/1006.2788}{{\tt arXiv:1006.2788}}.

\bibitem{Gaiotto:2010fk}
D.~Gaiotto, J.~Maldacena, A.~Sever, and P.~Vieira, ``{Bootstrapping null
  polygon Wilson loops},''
  \href{http://dx.doi.org/10.1007/JHEP03(2011)092}{{\em JHEP} {\bf 1103} (2011)
   092},
\href{http://arxiv.org/abs/1010.5009}{{\tt arXiv:1010.5009}}.

\bibitem{Basso:2010in}
B.~Basso, ``{Exciting the GKP string at any coupling},''
  \href{http://dx.doi.org/10.1016/j.nuclphysb.2011.12.010}{{\em Nucl.Phys.}
  {\bf B857} (2012)  254--334},
\href{http://arxiv.org/abs/1010.5237}{{\tt arXiv:1010.5237}}.

\bibitem{Basso:2011rs}
B.~Basso, ``{An exact slope for $AdS$/CFT},''
\href{http://arxiv.org/abs/1109.3154}{{\tt arXiv:1109.3154}}.

\bibitem{Gromov:2009tq}
N.~Gromov, ``{$Y$-system and quasi-classical strings},''
  \href{http://dx.doi.org/10.1007/JHEP01(2010)112}{{\em JHEP} {\bf 1001} (2010)
   112},
\href{http://arxiv.org/abs/0910.3608}{{\tt arXiv:0910.3608}}.

\bibitem{abjm}
O.~Aharony, O.~Bergman, D.~L. Jafferis, and J.~Maldacena, ``{${\cal N}=6$
  superconformal Chern-Simons-matter theories, M2-branes and their gravity
  duals},'' \href{http://dx.doi.org/10.1088/1126-6708/2008/10/091}{{\em JHEP}
  {\bf 10} (2008)  091},
\href{http://arxiv.org/abs/0806.1218}{{\tt arXiv:0806.1218}}.

\bibitem{dt}
N.~Drukker and D.~Trancanelli, ``{A supermatrix model for ${\cal N}=6$ super
  Chern-Simons-matter theory},''
  \href{http://dx.doi.org/10.1007/JHEP02(2010)058}{{\em JHEP} {\bf 02} (2010)
  058},
\href{http://arxiv.org/abs/0912.3006}{{\tt arXiv:0912.3006}}.

\bibitem{Beisert:2005tm}
N.~Beisert, ``{The $SU(2|2)$ dynamic S-matrix},'' {\em Adv.Theor.Math.Phys.}
  {\bf 12} (2008)  945, \href{http://arxiv.org/abs/hep-th/0511082}{{\tt
  hep-th/0511082}}.

\bibitem{Beisert:2006qh}
N.~Beisert, ``{The analytic Bethe ansatz for a chain with centrally extended
  $su(2|2)$ symmetry},''
  \href{http://dx.doi.org/10.1088/1742-5468/2007/01/P01017}{{\em J.Stat.Mech.}
  {\bf 0701} (2007)  P01017}, \href{http://arxiv.org/abs/nlin/0610017}{{\tt
  nlin/0610017}}.

\bibitem{Beisert:2006ib}
N.~Beisert, R.~Hernandez, and E.~Lopez, ``{A crossing-symmetric phase for
  $AdS_5 \times S^5$ strings},''
  \href{http://dx.doi.org/10.1088/1126-6708/2006/11/070}{{\em JHEP} {\bf 0611}
  (2006)  070},
\href{http://arxiv.org/abs/hep-th/0609044}{{\tt hep-th/0609044}}.

\bibitem{Vieira:2010kb}
P.~Vieira and D.~Volin, ``{Review of $AdS$/CFT integrability, Chapter III.3:
  The dressing factor},''
  \href{http://dx.doi.org/10.1007/s11005-011-0482-0}{{\em Lett.Math.Phys.} {\bf
  99} (2012)  231--253},
\href{http://arxiv.org/abs/1012.3992}{{\tt arXiv:1012.3992}}.

\end{thebibliography}\endgroup
\end{document}